\documentclass{jfm} 
\usepackage{graphicx}
\usepackage{comment}
\usepackage{amsmath,amssymb}
\usepackage{color}
\usepackage{bm}

\usepackage{xcolor}
\usepackage{hyperref}
\newcounter{aqctr}
\newenvironment{author-query}
{\refstepcounter{aqctr}\par\vspace{\baselineskip}\noindent
\color{red}\textbf{Author Query/Comment AQ \arabic{aqctr}.}}
{\par\vspace{\baselineskip}\normalcolor}

\newcommand{\rmi}{{\rm i}} 
\renewcommand{\d}{{\rm d}} 

\newcommand{\Pmt}{P_{-\frac{1}{2} + \rmi\tau}}
\shorttitle{Liquid jets with arbitrary contact angles}
\shortauthor{H. Miyoshi, H. Watanabe, I. Kikuchi, and Y. Tagawa}

\title{Impulse-induced liquid jets from bubbles with arbitrary contact angles}

\author{Hiroyuki Miyoshi\aff{1}
\corresp{\email{hiroyukimiyoshi@g.ecc.u-tokyo.ac.jp}},
  Hiroya Watanabe\aff{2}, Ishin Kikuchi\aff{2}, and Yoshiyuki Tagawa\aff{2}}

\affiliation{\aff{1}Department of Mathematical Engineering and Information Physics, The University of Tokyo, 7-3-1, Hongo, Bunkyo-ku, Tokyo, Japan 
\aff{2}Department of Mechanical Systems Engineering, Tokyo University of Agriculture and Technology, 2-24-16, Naka-cho, Koganei, Tokyo, 184-8588, Japan }

\begin{document}

\maketitle


\begin{abstract} 
This paper investigates the relationship between the contact angle of a spherical bubble attached to a tube submerged in a container and the jet speed induced by an impulsive acceleration at its base. 
While it has been well established that bubble geometry strongly influences the ejection speeds of liquid jets, 
mathematical studies of liquid jets with arbitrary bubble shapes remain limited. 
In this work, we derive a pressure impulse in the small-cavity limit as a tractable integral of classical Legendre functions. 
It is shown that the jet speed can be divided into two components: (i) the velocity induced by the hydrostatic pressure impulse distribution created by the curvature of the bubble, and (ii) the velocity induced by the distribution of the submersion of the tube in a container. 
This decomposition reveals that an optimal bubble curvature emerges only when the tube is submerged: the optimality is absent for non-submerged configurations, where the jet speed increases monotonically with bubble depth. 
Experiments confirm this non-monotonicity and quantitatively support the predicted shift of the optimal geometry with submersion depth.
\end{abstract}


\section{Introduction}




Liquid jets underpin a wide range of engineering applications, 
from precision manufacturing to high-resolution inkjet technologies~\citep{Lohse2022}. 
Eggers and Villermaux provide a comprehensive synthesis of the governing physics and canonical models in their seminal review~\citep{Eggers2008-of}, which continues to guide both fundamental and applied studies of liquid jets.
The same physical principles have advanced industrial applications such as needle-free drug delivery~\citep{baxter2006needle,srivastava2025detailed, mohizin2023}, additive manufacturing, and microfluidic actuation. 
Engineering applications such as tissue incision, metal cutting, and precision cleaning demand rapidly accelerating jets that can be produced without elaborate apparatus~\citep{tagawa2012highly}.


Liquid jets occur as a result of an impulse in a container with a solid boundary, 
where a cavity or object exists in the container. 
For instance, \citet{bartolo2006} studied jet bursts due to the cavity collapse of droplets deformed by impulses between the droplets and hydrophobic surfaces. 
Several studies have modelled the dynamics of such high-speed jets. 
\citet{gekle2009high} formulated an axisymmetric potential model that links cavity collapse to jet velocity. 
\citet{tagawa2012highly} examined the conditions required to form highly focused jets, while 
Rohilla {\em et al.} investigated how geometric parameters affect the dynamics of jets by using electrical sparks to produce bubble cavitation~\citep{rohilla2023}. 
\citet{Antkowiak2007-dk} derived a semi-analytical solution for the velocity ejected from a spherical cavity after impulsive forcing under mixed boundary conditions. More recently, \citet{dixit2025viscoelastic} showed that incorporating polymers into the liquid within spherical cavities can sustain even higher jet speeds by leveraging viscoelastic stresses. 
When the acceleration of a liquid occurs within a short time, 
it is useful to consider the limiting case of an impulsive change~\citep{batchelor2000introduction}. 
In such scenarios, a pressure impulse induced by an impulse satisfying the Laplace equation is used to investigate the dynamics of liquid jets. 
\citet{philippi2018} showed that the flow induced by such a pressure impulse reduces to the solution of a Laplace problem with Dirichlet and Neumann boundary conditions in hemispherical fluid regions. 
Recent experiments by \citet{Kurihara2025} further demonstrated that, under short-time acceleration, pressure fluctuations inside the liquid are strongly localized in time and can be quantitatively characterized within the pressure--impulse framework, providing experimental support for the applicability of such modelling to impact-driven jet formation. 

The mathematical modelling of Antkowiak's work has been validated through liquid jet experiments.
\citet{Onuki2018-jx} invented an apparatus for generating highly-focused liquid jets using a submerged gas--liquid interface inside a tube. They showed that the submerged liquid dramatically increases the speed of jets and that the speed increases as the level of submersion increases. 
By refining the jet apparatus used by Onuki {\em et al.}, \citet{kamamoto2021} and \citet{kobayashi2024} developed a novel coating technology with highly viscous jets. 
\citet{Kiyama2016-mi} conducted a series of experiments to identify important factors affecting the liquid jet speed. 
\citet{watanabe2025} compared full numerical results for a pressure impulse in a tapered container with results obtained using a simplified model. 
It was determined that the jet velocity depends on both the total flow rate due to changes in the cross-sectional area of the container and changes in the pressure impulse along the centre line. 
It has also been reported that, with regard to the experiment by Antkowiak {\em et al.,} the bounce height due to the pressure impulse in a rotational container is dramatically lower than that in an irrotational container~\citep{andrade2023, xie2025}.


It is also widely known that the shape of a spherical cavity at the impulse time significantly affects the velocity of jets from the cavity~\citep{Gordillo2023-wo}. 
The geometry of the liquid container also significantly affects jet speed and jet stability. 
\citet{reuter2021supersonic} studied the supersonic jets from conical liquid containers 
and found that the jet speed increases as the cone angle increases. 
Krishnan {\em et al.} found that the jet velocity scales with the geometric mean
of two physical velocities: the impact
velocity and the meniscus-deformation velocity scale~\citep{krishnan2022}. 
Cheng {\em et al.} studied impacts on free surfaces to investigate viscous effects on focused jet formation~\citep{cheng2024}. 
Although there have been a number of experimental studies on the relationship between jet velocity and meniscus shape, as summarized above, such mathematical modelling remains limited. 

In this paper, we address this gap by developing an analytical pressure--impulse solution for impulsively generated liquid jets from spherical bubbles with arbitrary contact angles. 
Using toroidal coordinates, originally introduced for solving Dirichlet boundary value problems by~\cite{lebedev1965special}, together with special function representations based on Legendre functions, we derive closed-form expressions for the pressure impulse and the resulting jet velocity in the small-bubble limit. 
Related formulations using toroidal coordinates have recently been applied to problems such as bubble evaporation~\citep{popov2005evaporative,wilson2023} and the interaction of multiple bubbles on substrates~\citep{Masoud2021-ke,DOMBROVSKY2025103605}, demonstrating their versatility in free boundary problems.

A key outcome of the present analysis is that the jet velocity can be decomposed into two physically distinct contributions:
(i) a curvature-induced focusing term associated with the bubble geometry, and
(ii) a submersion-induced term arising from the pressure--impulse redistribution imposed by the surrounding container.
This decomposition reveals an optimal bubble shape that maximizes the jet velocity when the tube is submerged. This effect does not arise in the absence of submersion, where the jet speed increases monotonically with bubble depth. 
Image-based impact experiments validate the theoretical predictions and confirm that the optimal geometry emerges from the competition between these two mechanisms.

This new solution provides a hybrid approach that accounts for the container boundary. 
The toroidal coordinates and cylindrical coordinates are mixed using a change of variables, and the solution is given by a series expansion of the basis obtained from even-number derivatives of the analytical solution derived in Section~\ref{sec:small}.  
A key point that is often missed in intuitive interpretations is that submersion does not simply shift the hydrostatic impulse by an additive constant.
Because the jet speed is set by the gradient of the pressure impulse, the submersion depth enters the velocity through a distinct harmonic component associated with the boundary condition at the free surface. This yields a natural decomposition of the jet speed into two contributions, which serves as the central organising principle of this paper.

Numerical results for jet speed as a function of angle show that there is an optimal angle that maximizes the jet speed. This trend is validated in an actual experiment in which impacts on the bottom of a container of liquid are induced. Most importantly, the critical angle of a spherical bubble decreases when the tube is submerged in a liquid. This result indicates that the control of the shape of the spherical cavity is important for producing high-speed jets. 

\begin{figure}
\centerline{\includegraphics[width=0.89\linewidth]{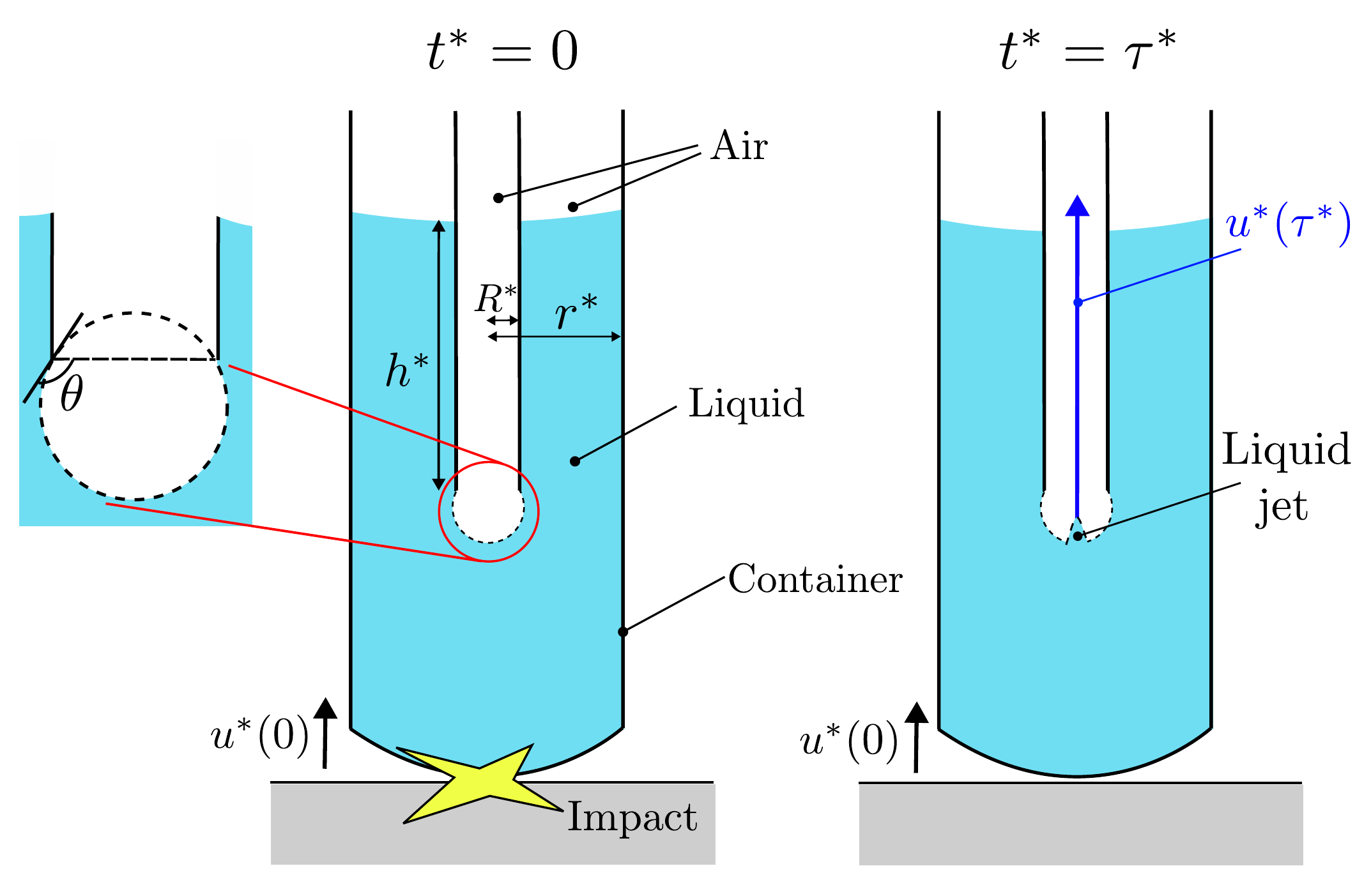}} 
    \caption{A liquid jet created by an impact on the ground. 
    At the initial time $t^*=0$, a falling container experiences an impact on its bottom, which produces a variation of the pressure impulse $\Pi^*$ in a cylindrical container. 
    The bubble region on the tip of the tube is assumed to be spherical, which is consistent with some experimental results (see~\cite{Onuki2018-jx}). After the impact, a liquid jet is generated by the non-uniform distribution of the pressure impulse. }
    \label{fig:abst}
\end{figure}

The rest of the paper is organized as follows. 
Section~\ref{sec:probform} formulates the problem after the impulse and derives a 3D axisymmetrical Laplace equation for a pressure impulse with mixed boundary conditions on the meniscus and the wall. 
Using toroidal coordinates, analytical formulas for the pressure impulse with a small-bubble limit are obtained in Section~\ref{sec:small}. 
For a general region with a bubble, a superposition of the solution derived in Section~\ref{sec:small} with some derivatives is used to obtain the solution for the Laplace equation in Section~\ref{sec:exact}. 
Finally, the experimental results are presented in Section~\ref{sec:experiment}, and the conclusions are given in Section~\ref{sec:conclusion}. 


\section{Problem formulation} \label{sec:probform}
Throughout this paper, we use an asterisk to represent dimensional quantities. 
For non-dimensional quantities, we use normal letters. 
Consider a spherical bubble attached to the tip of a tube, where the tube is submerged to a depth $h^*$ in a rigid container outside the tube. The container suddenly experiences an impulse at its bottom due to an impact, and the jet bursts from the bottom of the controlled spherical bubble. The angle between the meniscus of the bubble and the horizontal line is defined as $\theta$. The radius of the cylindrical tube inside the container is $R^*$, and that of the container is $r^*$. Figure~\ref{fig:abst} shows a schematic of the geometric parameters. 

We define the velocity before and after the impact as $\bm{u}^*(0)$ and $\bm{u}^*(\tau^*)$, where $\tau^*$ is an impact duration period. 
According to a classical result from~\cite{batchelor2000introduction} and recent work on liquid jets by~\cite{Antkowiak2007-dk}, it is assumed that the fluid is inviscid and irrotational during the transient time of the impulse. The velocity of the liquid $\bm{u}^*$ is then governed by  
\begin{align}
    \frac{\partial \bm{u}^*}{\partial t^*} = -\frac{1}{\rho^*}\nabla^* 
    p^*,\label{eq:dudt}
\end{align}
where $\nabla^*$ is a dimensional gradient operator, $p^*$ is a fluid pressure, and $\rho^*$ is the density of the liquid. 
By integrating~(\ref{eq:dudt}) with respect to the short time period of the impact $\tau^*$, the velocity $\bm{u}^*(\tau^*)$ satisfies 
\begin{align}
    \bm{u}^*(\tau^*) - \bm{u}^*(0)= -\frac{1}{\rho^*} \nabla^* \Pi^*. \label{eq:Pstar_deriv}
\end{align}
Here $\Pi^*$ is the pressure impulse defined by the integral of the pressure $p^*$ over the short impact duration time $\tau^*$. 
Since $\bm{u}^*$ is assumed to be incompressible, the pressure impulse $\Pi^*$ satisfies the 3D Laplace equation in the region filled with liquid; that is, 
\begin{align}
    \nabla^{*2} \Pi^* = 0
\end{align}
in the cylindrical domain outside the bubble. 
The jet velocity $\bm{u}^*(\tau^*)$ is calculated using the spatial derivative of $\Pi^*$, as given in~(\ref{eq:Pstar_deriv}) after obtaining $\Pi^*$. 

\begin{figure}
\centerline{\includegraphics[width=0.99\linewidth]{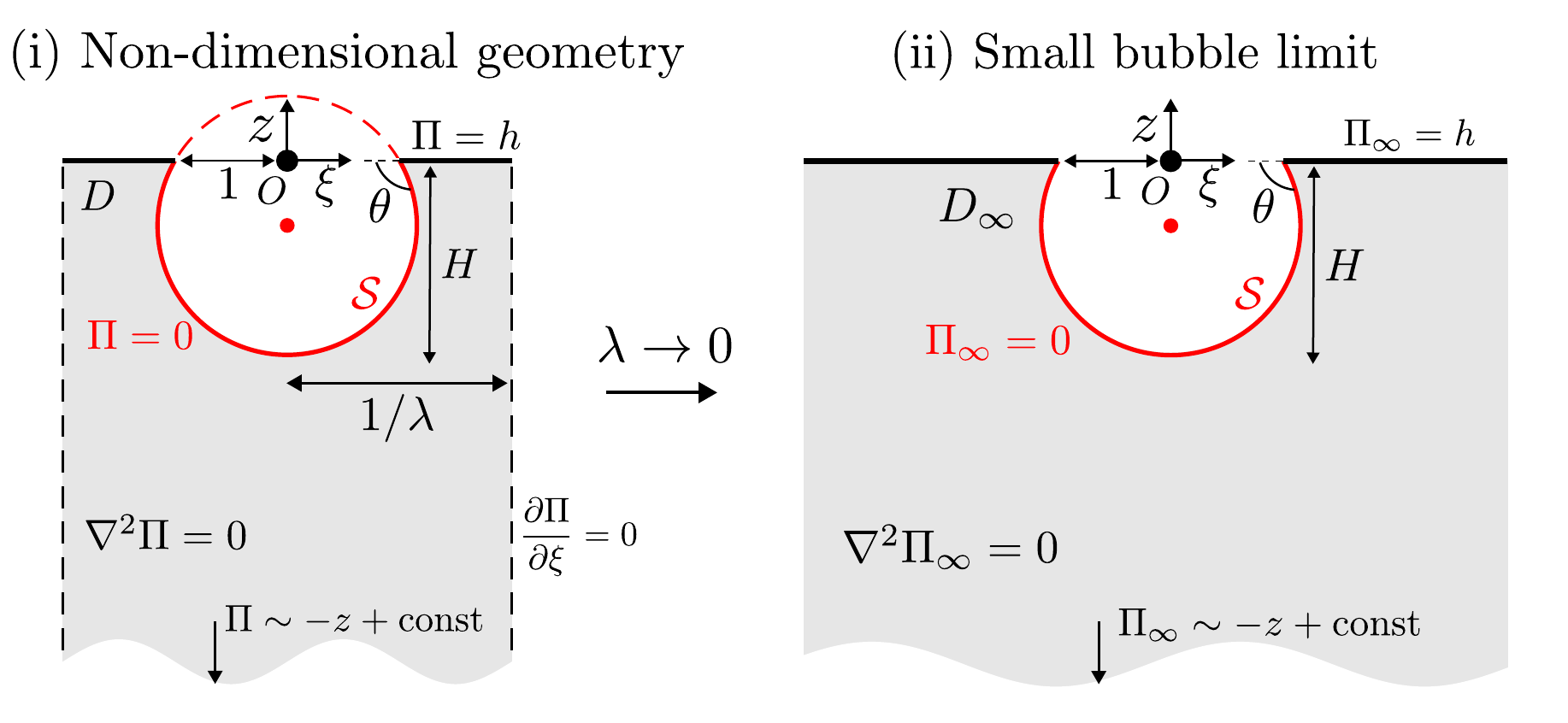}} 
    \caption{(i) The non-dimensional boundary value problem for a harmonic pressure impulse $\Pi$ and (ii) the geometry $D_\infty$ in the small-bubble limit. When $\lambda\rightarrow 0$, the geometry is a semi-infinite half 3D space outside a spherical-shaped bubble, denoted by $D_\infty$. Note that the value $H$ is related to $\theta$ as $H = \tan(\theta/2)$. The impulse $\Pi$ is approximated well by $\Pi_\infty$ as $\lambda \rightarrow 0$.  }
    \label{fig:forpaper}
\end{figure}

It is convenient to non-dimensionalize the pressure impulse and the coordinates, that is, $\Pi^* = \Pi \cdot \rho^* U^*_0 R^*$ and $(x^*,y^*,z^*) = (R^*x,R^*y,R^* z)$. With this scaling, the radius of the inner tube is always set to $1$. $U_0^*$ is the velocity of the container in the $z^*$-direction at the initial time due to the elastic collision at the bottom. Furthermore, the non-dimensional geometric parameters are defined by $h:=h^*/R^*$, $\lambda:=R^*/r^* $. 
The non-dimensional geometry is shown on the left of Figure~\ref{fig:forpaper}. 

Assuming axial symmetry of the shape of the cavity and container, 
it is convenient to study the jet speed in a cylindrical coordinate system defined by $(\xi,z)$. 
We consider a general solution to the 3D potential problem in a cylindrical region outside of a spherical cavity. The boundary of a spherical bubble with a contact angle $\theta$ is defined as $\mathcal{S}$:
\begin{align}
    \mathcal{S} := \left\{(\xi,z)\Big{|} \ \xi^2 + \left(z - \frac{1}{\tan\theta} \right)^2 = \frac{1}{\sin^2 \theta},\  z\leq0 \right\}.
\end{align}
Note that the depth of a bubble defined by $H$ is related to the contact angle $\theta$ by 
\begin{align}
    H  = \tan(\theta/2). \label{eq:H_theta}
\end{align}

We assume that the container length in the $z$-direction is much greater than the radius of the tube and the container. Hence, the elastic collision on the bottom gives a uniform velocity distribution at $z\rightarrow -\infty$ after the impact.  

The boundary condition for the interface is the Euler--Bernoulli equation at the free interface, which reduces to $\Pi=0$ on the meniscus $\mathcal{S}$, and on the top of the liquid interface in the container at $1<\xi<1/\lambda$, $z=h$~\citep{gordillo2020impulsive,Onuki2018-jx}. 
Furthermore, assuming that the uniform gradient of the pressure impulse appears in the region above the bubble region, it is possible to approximate that the pressure impulse satisfies $\Pi=h$ on $z=0$, $1<\xi<1/\lambda$. This approximation was also adopted in previous work by~\cite{Onuki2018-jx}, where it was shown to be in good agreement with experimental results for predicting jet velocities. 
For the side of the container, a no-penetrating condition is imposed; that is, $\partial \Pi/\partial \xi= 0$ for $\xi = 1/\lambda$, $z<0$, as the movement of the container is assumed to be only along the $z$-axis. 
As a result, the non-dimensional pressure impulse $\Pi$ satisfies the following mixed boundary value problem in cylindrical coordinates: 
\begin{align}
\left\{
\begin{aligned}
    \textrm{(i)}\quad &\nabla^{2} \Pi = 0,\quad &&(\xi,z) \in D,\\
    \textrm{(ii)}\quad &\frac{\partial \Pi}{\partial \xi}(1/\lambda,z) = 0,\quad &&z<0\\
    \textrm{(iii)}\quad &\Pi(\xi,z)=0,\quad &&(\xi,z)\in \mathcal{S},\\
    \textrm{(iv)}\quad &\Pi(\xi,0) =h,\quad &&1\leq \xi \leq 1/\lambda,\\
    \textrm{(v)}\quad & \Pi(\xi,z) = -z + {\rm const},\quad &&z \rightarrow -\infty,
    \end{aligned}\label{eq:problem}
\right.
\end{align}
where the region $D$ is the fluid region for $z<0$, as shown in the left of Figure~\ref{fig:forpaper}. 
After obtaining $\Pi$, the initial jet speed is calculated using the velocity at the bottom of the cavity with respect to the moving frame with Eq.~(\ref{eq:Pstar_deriv}).

\section{The small-bubble limit}\label{sec:small}
This section studies the small-bubble limit, that is, when the cavity is sufficiently small compared to the radius of the container shown in the right of Figure~\ref{fig:forpaper}. 
This case corresponds to the ratio of the radius of the inner tube to that of the container tending to $0$, that is, $\lambda\rightarrow 0$. 
The region is now a semi-infinite lower-half 3D region outside the spherical bubble $\mathcal{S}$, denoted by $D_\infty$. 
For this limiting case, it is possible to obtain an analytical solution for the field $\Pi_\infty$ and thus the jet velocity induced by the impulsive acceleration. 

As $\lambda \rightarrow 0$, the no-penetrating condition on the side wall in (ii) in~(\ref{eq:problem}) vanishes. The pressure impulse defined by $\Pi_{\infty}(\xi,z)$ now satisfies 
\begin{align}
\left\{
\begin{aligned}
    \textrm{(i)}\quad &\nabla^2 \Pi_{\infty} = 0,\quad (\xi,z) \in D_\infty,\\
    \textrm{(iii)}\quad &\Pi_{\infty}(\xi,z)=0,\quad (\xi,z)\in \mathcal{S},\\
    \textrm{(iv)}\quad &\Pi_{\infty}(\xi,0) =h,\quad 1\leq \xi \leq \infty,\\
    \textrm{(v)}\quad & \Pi_{\infty}(\xi,z) = - z + {\rm const},\quad z \rightarrow -\infty.
\end{aligned}
\right.\label{eq:F_problem}
\end{align}

\begin{figure}
\centerline{\includegraphics[width=0.69\linewidth]{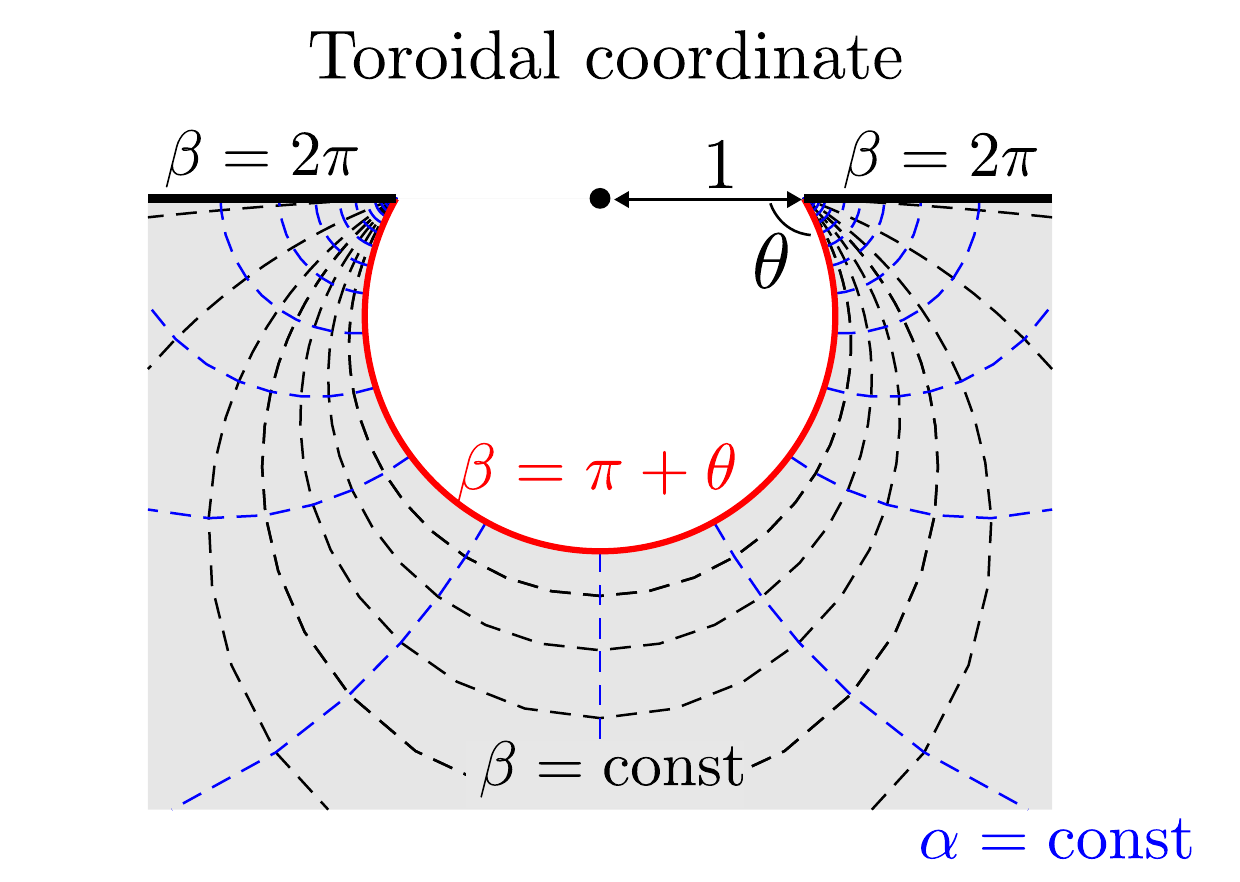}} 
    \caption{Toroidal coordinates $(\alpha,\beta)$ for solving the mixed boundary value problem for a pressure impulse $\Pi_\infty$. Cylindrical coordinates $(\xi,z)$ are mapped to the new coordinates $(\alpha,\beta)$ using the transformation in~(\ref{eq:change_xiz}). The boundary of the bubble is expressed by $\beta=\pi+\theta$. }
    \label{fig:toroidal}
\end{figure}

To solve the boundary value problem~(\ref{eq:F_problem}), it is convenient to change the cylindrical coordinates $(\xi,z)$ to toroidal coordinates $(\alpha,\beta)$~\citep{lebedev1965special}, where all boundaries of $D_\infty$ can be expressed by a contour line in this coordinate system. 
Following a mathematical setting provided in Chapter 8 of~\cite{lebedev1965special}, consider a change of variables as follows:
\begin{align}
      \xi(\alpha,\beta)=\frac{\sinh\alpha}{\cosh\alpha-\cos\beta},\quad z(\alpha,\beta)=\frac{\sin\beta}{\cosh\alpha-\cos\beta}, \label{eq:change_xiz}
\end{align}
where the parameter spaces for $\alpha$ and $\beta$ in the domain $D_\infty$ are $0 < \alpha < \infty,\  \pi+\theta < \beta < 2\pi$.
The inverse transform is given by 
\begin{align}
    \alpha(\xi,z) = \frac{1}{2}\log\left(\frac{z^2 + (\xi + 1)^2}{z^2 + (\xi - 1)^2} \right),\quad \beta(\xi,z)= \pi+\tan^{-1}\left(\frac{2z}{\xi^2 + z^2 - 1} \right),
\end{align}
where $\beta$ is chosen so that $\pi +\theta < \beta < 2\pi$. 
Using this coordinate system, the boundary of a spherical bubble $\mathcal{S}$ and the plane $1<\xi$ with $z=0$ correspond to $\beta=\pi + \theta$ and $\beta = 2\pi$, respectively. 
The correspondence between the geometry and parameters $\alpha$ and $\beta$ is shown in Figure~\ref{fig:toroidal}. 
Using the new coordinates and the mathematical formulation in~\cite{lebedev1965special}, the solution $\Pi_\infty(\xi,z)$ for the boundary value problem~(\ref{eq:F_problem}) is given by  
\begin{align}
    \Pi_{\infty}(\xi,z) = \Pi_f(\xi,z)+ h \Pi_g(\xi,z), \label{eq:Pi_infty}
\end{align}
where 
\begin{align}
    &\Pi_f(\xi,z) := -z - f_\theta(\alpha(\xi,z),\beta(\xi,z)),\\
    &\Pi_g(\xi,z) := 1- g_\theta(\alpha(\xi,z),\beta(\xi,z),
\end{align}
and 
\begin{align}
    f_\theta(\alpha,\beta) &:= 2\sqrt{2(\cosh\alpha - \cos\beta)}\int_0^\infty \tau \frac{\sinh \theta \tau \sinh(2\pi-\beta)\tau }{\cosh \pi\tau \sinh (\pi-\theta)\tau }P_{-\frac{1}{2} + \rmi\tau}(\cosh\alpha)\d \tau,
    \label{eq:pressure_asymp1} \\
    g_\theta(\alpha,\beta) &:= \sqrt{2(\cosh\alpha - \cos\beta)}\int_0^\infty \frac{\cosh \theta \tau \sinh(2\pi-\beta)\tau }{\cosh \pi\tau \sinh (\pi-\theta)\tau }P_{-\frac{1}{2} + \rmi\tau}(\cosh\alpha)\d \tau. \label{eq:pressure_asymp2}
\end{align}
The cylindrical coordinates $(\xi,z)$ are related to $(\alpha,\beta)$ through the coordinate transformation~(\ref{eq:change_xiz}). The function $P_\nu(z)$ is a special function called a Legendre function of the first kind with degree $\nu$. 
The derivation of this formula is provided in detail in Appendix~\ref{app:1}. 

Note that $\Pi_f(\xi,z)$ is a pressure impulse induced by the curvature of the spherical bubble with no submersion, and $\Pi_g$ is a pressure impulse induced by the submersion of the tube. The total pressure impulse $\Pi_\infty$ is decomposed into $\Pi_f$ and $\Pi_g$ times the submersion length $h$. 

Although the expressions~(\ref{eq:pressure_asymp1}) and (\ref{eq:pressure_asymp2}) are similar to (8.12.12) in \cite{lebedev1965special}, these solutions are novel due to the differing boundary conditions (i), (iii)--(v). This is also one of the main contributions of this paper. 

This formula includes an integral of the function $\Pmt(\cosh\alpha)$ with respect to $\tau$. It is possible to calculate the function using some well-known integral representations~\citep{lebedev1965special}. Numerical computations for this function in our paper are given in Appendix~\ref{app:comp_legendre}.

The speed of the jet is calculated directly using the derivative of $\Pi_\infty$ in~(\ref{eq:Pi_infty}) with respect to $z$ evaluated at the bottom of the bubble. To achieve this, the derivatives of $f_\theta$ and $g_\theta$ are calculated using the chain rule as follows:
\begin{align}
    \frac{\partial f_\theta}{\partial z} = \frac{\partial\alpha}{\partial z}\frac{\partial f_\theta}{\partial \alpha} + \frac{\partial\beta}{\partial z}\frac{\partial f_\theta}{\partial \beta},\quad \frac{\partial g_\theta}{\partial z} = \frac{\partial\alpha}{\partial z}\frac{\partial g_\theta}{\partial \alpha} + \frac{\partial\beta}{\partial z}\frac{\partial g_\theta}{\partial \beta}. \label{eq:dfdz_dgdz}
\end{align}
The jet speed is evaluated at the bottom of the spherical bubble located at $(\xi,z)=(0,-H)$. The parameters there are $\alpha=0$ and $\beta = \pi + \theta$. Note that the derivatives of $f_\theta$ and $g_\theta$ with respect to $\alpha$ and $\beta$ in (\ref{eq:dfdz_dgdz}) can be explicitly evaluated. For example, the derivative of $g_\theta$ with respect to $\beta$ is given by 
\begin{align}
        \frac{\partial g_\theta}{\partial \beta} &= \frac{\sin \beta}{ \sqrt{2(\cosh\alpha - \cos\beta)}}\int_{0}^\infty  \frac{\cosh \theta \tau \sinh(2\pi-\beta)\tau }{\cosh \pi\tau \sinh (\pi-\theta)\tau }P_{-\frac{1}{2} + \rmi\tau}(\cosh\alpha)\d \tau\nonumber\\
    &-\sqrt{2(\cosh\alpha - \cos\beta)}\int_0^\infty \tau \frac{\cosh \theta \tau \cosh(2\pi-\beta)\tau }{\cosh \pi\tau \sinh (\pi-\theta)\tau }P_{-\frac{1}{2} + \rmi\tau}(\cosh\alpha)\d \tau.\label{eq:partialGpartialbeta}
\end{align}
By using the Mehler--Fock theorem and the integral representation of a Legendre function~\citep{lebedev1965special}, the speed at the bottom of the cavity $v_f(\theta)$ in a frame moving with the tube is given explicitly by
\begin{align}
    v_f(\theta) :=  \frac{\partial f_\theta}{\partial z}(0,\pi+\theta) = \sin^2 \left(\frac{\theta}{2} \right) + 8 \cos^3\left(\frac{\theta}{2}\right)\int_0^{\infty} \tau^2 \frac{\sinh\theta \tau \cosh(\pi-\theta)\tau}{\cosh\pi\tau \sinh(\pi-\theta)\tau}\d \tau, \label{eq:velocity1}
\end{align}
and
\begin{align}
    v_g(\theta):=\frac{\partial g_\theta}{\partial z}(0,\pi+\theta) =\frac{1}{2}\sin\theta + 4\cos^{3}\left(\frac{\theta}{2}\right)\int_0^\infty \tau \frac{\cosh \theta \tau \cosh(\pi-\theta)\tau }{\cosh \pi\tau \sinh (\pi-\theta)\tau }\d \tau. \label{eq:velocity2}
\end{align}
Based on the pressure impulse decomposition (\ref{eq:Pi_infty}), the total jet velocity $v(\theta)$ is then equal to 
\begin{align}
v(\theta) = v_f(\theta) + h v_g(\theta),
\end{align}
which means that the velocity is composed of the velocity $v_f(\theta)$ due to the pressure impulse of the impact with $h=0$ and the velocity $v_g(\theta)$ caused by the pressure distribution due to the submergence of the tube. 
This is natural since the solution~(\ref{eq:Pi_infty}) is composed of the hydrostatic pressure impulse due to the bubble $\Pi_f$ and the pressure due to the submergence of the tube, which is $\Pi_g$ times the height $h$. 
This decomposition is important because the jet speed $v(\theta)$ has a local maximum with respect to $h$ and the bubble height $H=\tan(\theta/2)$.

\begin{figure}
\centerline{\includegraphics[width=0.99\linewidth]{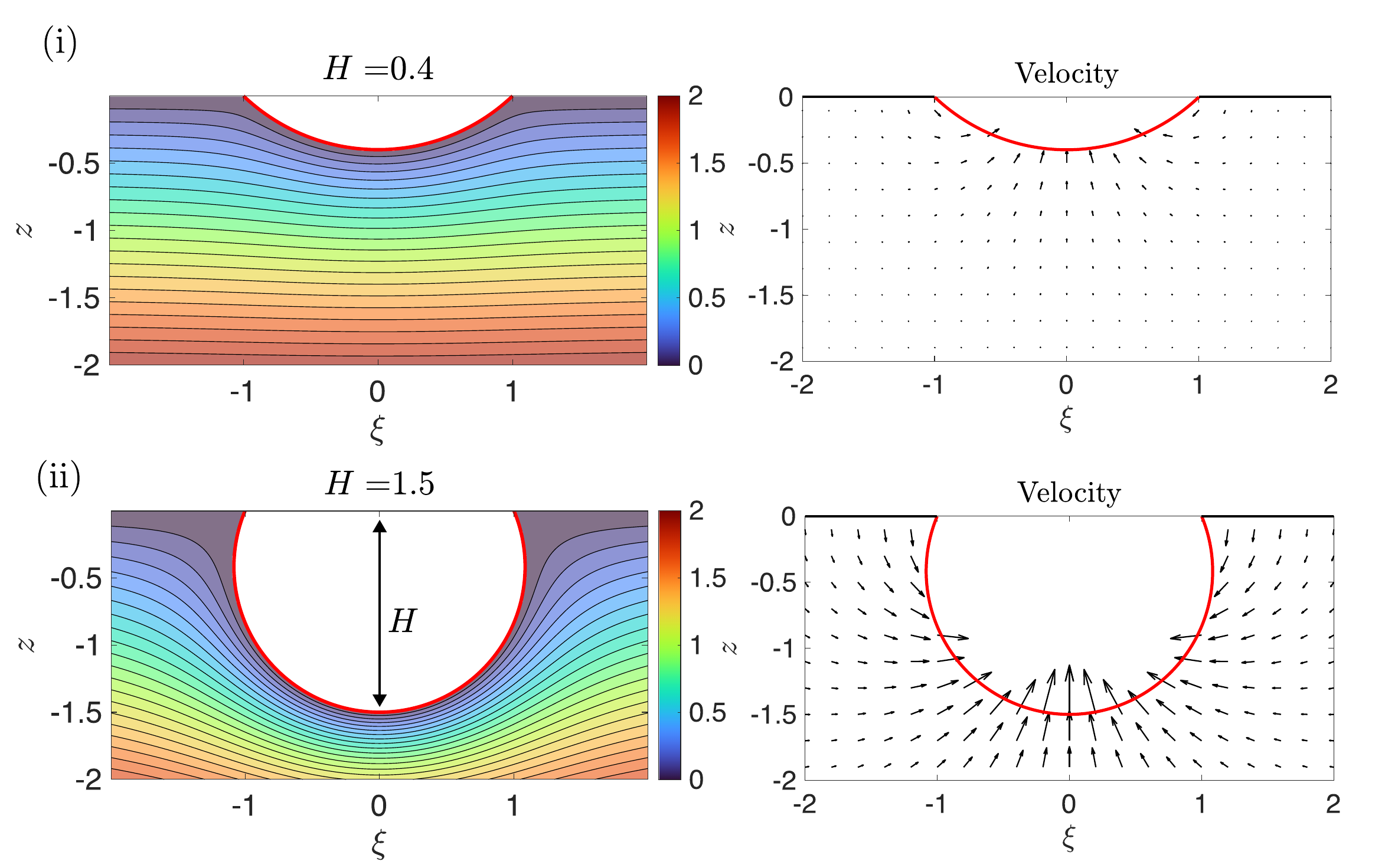}} 
    \caption{Contour plots for the harmonic pressure impulse $\Pi_\infty(\xi,z)$ in~(\ref{eq:pressure_asymp1}) and the velocity in~(\ref{eq:velocity1}). (i) $H=0.4$ and (ii) $H=1.5$. The value $H$ is related to the angle $\theta$ as $H  = \tan(\theta/2)$. }
    \label{fig:figure_vel_cont1}
\end{figure}  

\begin{figure}
\centerline{\includegraphics[width=0.99\linewidth]{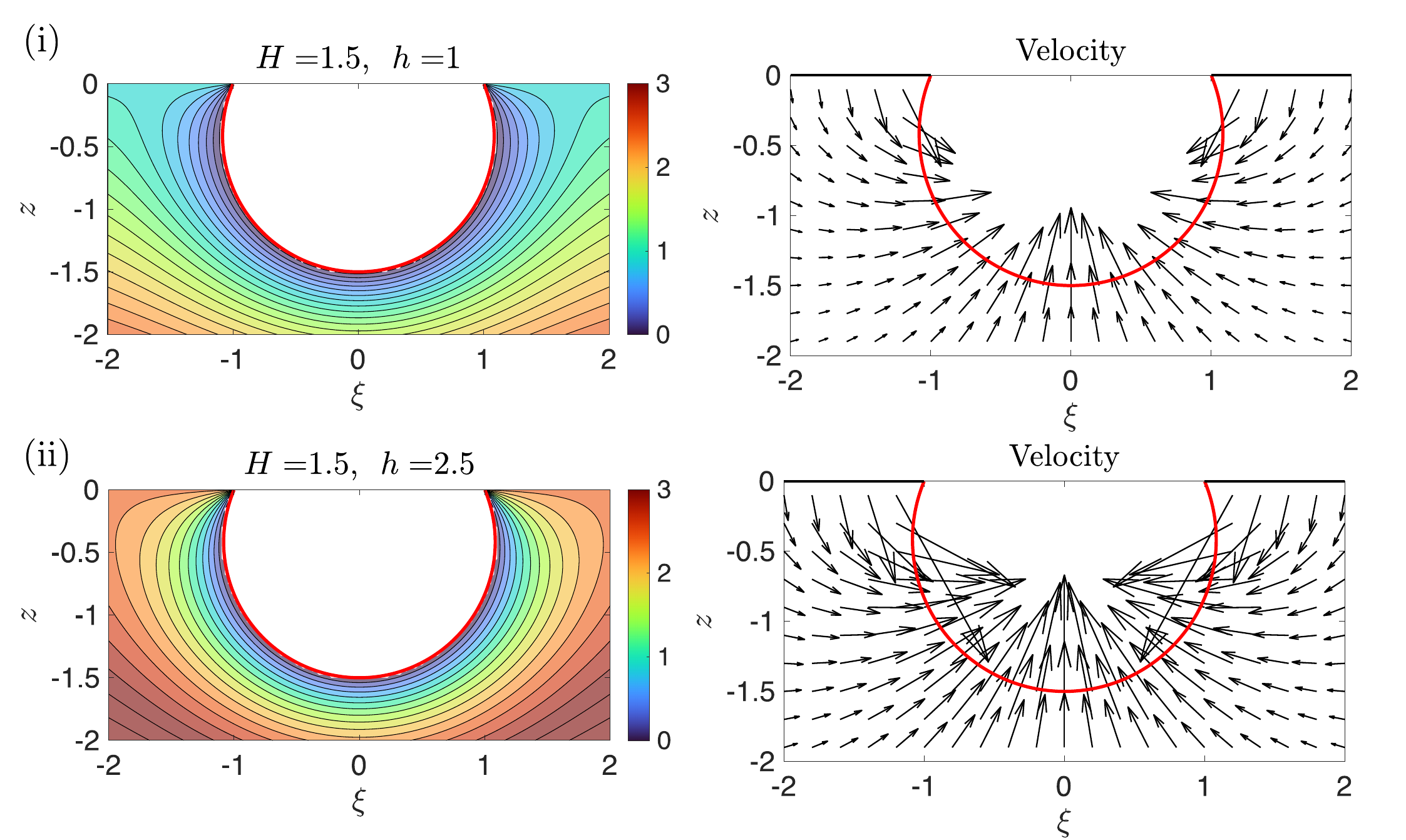}} 
    \caption{Contour plots for the harmonic pressure impulse $\Pi_\infty(\xi,z)$ in~(\ref{eq:pressure_asymp1}) and velocity in~(\ref{eq:velocity1}) for $H=1.5$. Note that $\Pi_\infty(\pm 1,0)$ is discontinuous due to the boundary conditions. (i) $h=1.0$ and (ii) $h=2.5$. The cavity height is fixed at $H=1.5$. }
    \label{fig:figure_vel_cont2}
\end{figure}  

The final integrals in~(\ref{eq:velocity1}) and (\ref{eq:velocity2}) do not appear to admit a simpler closed-form expression. However, since the integrand rapidly decays to zero as $\tau\rightarrow \infty$, they can be readily evaluated numerically. 
Nonetheless, it is possible to show that, for $\theta=\pi/2$, the integral expression~(\ref{eq:velocity1}) can be simplified as follows:
\begin{align}
    v_f(\pi/2) = \frac{1}{2} + 2\sqrt{2}\int_0^{\infty}\tau^2\frac{\cosh \frac{\pi}{2}\tau}{\cosh\pi\tau}\d \tau = 2.
\end{align}
This result corresponds to Figure 6 in~\cite{Antkowiak2007-dk} as $\lambda\rightarrow 0$, where the jet produced from hemispherical bubbles was studied. This velocity is consistent with the fact that the pressure impulse for $H=1$ ($\theta=\pi/2$) is explicitly written as $F_{\pi/2}(\xi,z) = -z - z/(\xi^2 + z^2)^{3/2}$. It is easy to see that the function satisfies $F_{\pi/2}(\xi,z)=0$ on the lower hemisphere and $z=0$, $\xi\geq 1$. This fact will be used in Section~\ref{sec:exact}. 
For $\theta=0$, it is easy to see that $v_f(0)=0$ and 
\begin{align}
    v_g(0) = 4\int_{0}^\infty \frac{\tau \d \tau}{\sinh \pi\tau} = 8\int_0^\infty \frac{\tau \d \tau}{e^{\pi\tau} - e^{-\pi\tau}} = 8 \lim_{\epsilon\rightarrow 0^+} \int_{\epsilon}^\infty \frac{\tau}{e^{\pi\tau}}\sum_{n=0}^\infty e^{-2n\pi\tau}\d \tau = 1.
\end{align}

Figure~\ref{fig:figure_vel_cont1} shows contour plots of $\Pi_f$ and velocity plots with cavity heights of (i) $H=0.4$ and (ii) $H=1.5$. Note that $H$ and $\theta$ are related to each other by the relation (\ref{eq:H_theta}). 
As the cavity height $H$ increases, the magnitude of the velocity increases. It can also be observed that the maximum velocity is attained at the bottom of the bubble. This is natural because the spherical bubble pushes as $H$ increases, and the interval between contour lines becomes small, so that the pressure gradient at the bottom becomes large. 
For $h>0$, the contour plots of $\Pi_\infty$ and velocity plots for a cavity height of $H=1.5$ with (i) $h=1$ and (ii) $h=2.5$ are shown in Figure~\ref{fig:figure_vel_cont2}. The velocity becomes large as $h$ increases. 

\begin{figure}
\centerline{\includegraphics[width=0.99\linewidth]{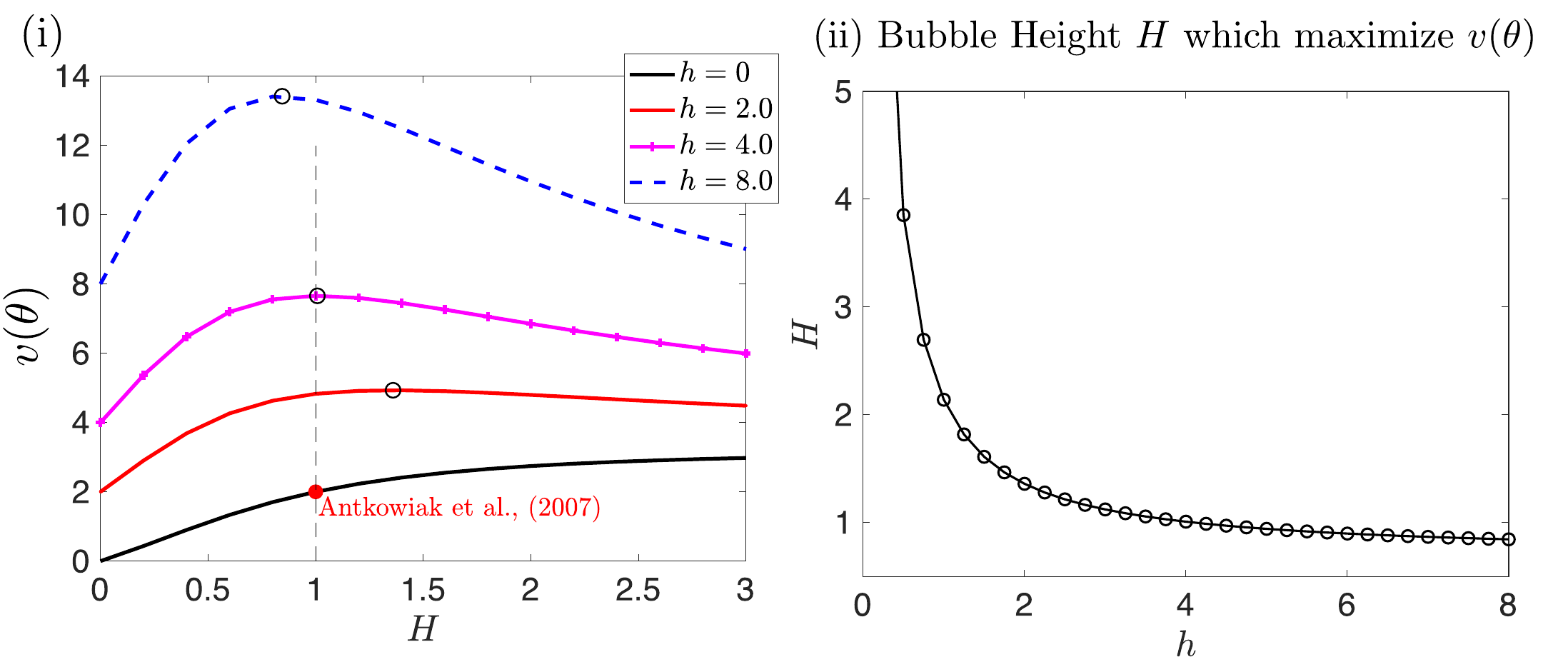}} 
    \caption{(i) The speed at the bottom of the bubble with different submersion depths $h$. 
    The jet velocity with four different $h$ are examined, that is, $h=0$, $h=2.0$, $h=4.0$, and $h=8.0$. The circle dots correspond to the points where the velocity $v(\theta)$ is maximized. (ii) The critical value for $H$ that maximizes the jet speed with respect to the depth $h$. }
    \label{fig:figure_vel_maximum}
\end{figure}

Figure~\ref{fig:figure_vel_maximum} shows the jet velocity $v(\theta)$ with respect to $H=\tan(\theta/2)$ for different submersion depths: $h=0$, $h=2.0$, $h=4.0$, and $h=8.0$. For $h=0$, the velocity monotonically increases as the height of the bubble $H$ increases. On the contrary, for $h>0$, there exists a critical value of $H$ that maximizes the velocity $v(\theta)$. 
It is also observed that the critical value of $H$ gets smaller as $h$ increases.


The formulas derived here work only when the walls and inner cavity are widely separated. When they are close, the jet velocity is not approximated well by the small-bubble limit. However, we will show that the formulas~(\ref{eq:pressure_asymp1}) and (\ref{eq:pressure_asymp2}) are in good agreement with a wide range of parameter spaces of $\lambda$ and $H$ in the next section.

\section{Exact solution for the pressure impulse using a series solution}\label{sec:exact}
Section~\ref{sec:small} presents an analytical formula for the pressure impulse field and velocity when the bubble size is small enough to neglect the side wall effect. 
Although analytical formulas for the fields and velocities of liquid jets have been presented, the presence of the wall affects both. 
\citet{Antkowiak2007-dk} studied the relationship between the jet velocity and changes in the cavity size and tube radius, where the contact angle of the cavity is always $\pi/2$. It was found, both numerically and experimentally, that the jet velocity decreases as the cavity approaches the wall boundary, that is, $\lambda\rightarrow 1$. It was also observed that the velocity in the $h=0$ case is approximately $1$. 

To study a similar situation to~\citet{Antkowiak2007-dk}, we present a semi-analytical approach for obtaining the pressure impulse in the presence of a wall boundary based on the analytical formulas in Section~\ref{sec:small}. Similar to the formulation in~\cite{Antkowiak2007-dk}, the field inside the tube outside the spherical bubble is given by the superposition of high-order fields derived from the pressure impulse of Eqs.~(\ref{eq:pressure_asymp1}) and~(\ref{eq:pressure_asymp2}). 

\subsection{Derivation of the semi-analytical solution for $\theta=\pi/2$ by Antkowiak {\em et al.} }
Antkowiak~{\em et al.} derived a semi-analytical solution for a potential problem in a tube outside a hemispherical bubble, that is, $\theta=\pi/2$~\citep{Antkowiak2007-dk}. 
\cite{Antkowiak2007-dk} started the formulation by considering the initial function: 
\begin{align}
    \phi_0(\xi,z) := \frac{z}{(\xi^2 + z^2)^{3/2}}.
\end{align}
It is easy to see that $\Pi(\xi,z):=-z - \phi_0(\xi,z)$ satisfies the conditions (i), (iii)--(v) in~(\ref{eq:problem}). To satisfy the boundary condition (ii), we consider the modified field $F_{0}(\xi,z) := \phi_0(\xi,z) + \psi_0(\xi,z)$, where $\psi_0(\xi,z)$ is added to satisfy the boundary condition (ii). By employing the condition $\psi_0(\xi,z)=0$ on $z=0$, $\psi_0(\xi,z)$ is represented by Fourier--sine functions with Bessel functions $I_0(\xi)$ as follows: 
\begin{align}
    \psi_0(\xi,z) := \int_{0}^\infty \eta_0(m)\sin(mz) I_0(m\xi) \d m, \label{eq:psi}
\end{align}
where $f_0(m)$ is calculated from the Fourier--sine transform of the function to satisfy the boundary condition on the side wall of the tube as follows:
\begin{align}
    \eta_0(m) = -\frac{2}{\pi}\int_{-\infty}^0 \frac{\sin m z}{m I_1(m/\lambda)} \frac{\partial \phi_0}{\partial \xi}(1/\lambda,z)\d z. \label{eq:f0_calc}
\end{align}
Although $-z - F_0(\xi,z)$ is not an exact solution for the boundary value problem with a tube as $-z-F_0(\xi,z)\neq 0$ on $\mathcal{S}$, it is possible to obtain the solution using a new basis related to $\phi_0(\xi,z)$. 
Consider an even-number derivative of $\phi_0(\xi,z)$ with respect to $z$:
\begin{align}
    \phi_{2n}(\xi,z) : = \frac{1}{(2n)!}\frac{\partial^{2n} \phi_0}{\partial z^{2n}}(\xi,z),\quad \psi_{2n}(\xi,z) :=\int_{0}^\infty \eta_{2n}(m)\sin(mz) I_0(m\xi) \d m, \label{eq:phi_2n}
\end{align}
where the coefficient $\eta_{2n}(m)$ is also calculated in a similar manner to~(\ref{eq:f0_calc}) by simply modifying $\phi_{0}(1/\lambda,z)$ as $\phi_{2n}(1/\lambda,z)$. The solution $\Pi(\xi,z)$ is the superposition of the functions $F_{2n}$ as follows:
\begin{align}
    \Pi(\xi,z) := -z - \sum_{n=0}^\infty A_{2n} F_{2n}(\xi,z),\quad F_{2n}(\xi,z) := \phi_{2n}(\xi,z) + \psi_{2n}(\xi,z). \label{eq:ant_semianaly}
\end{align}
Antkowiak {\em et al.} derived an explicit representation for $F_{2n}(m)$ and then a linear system for the coefficients $A_{2n}$, with a truncation order $N$. 

We show that this approach is applicable to problems that do not always have a hemispherical bubble. The only modification is the basis function $\phi_{2n}$.

\subsection{A semi-analytical solution for an arbitrary angle $\theta$}

\begin{figure}
\centerline{\includegraphics[width=0.99\linewidth]{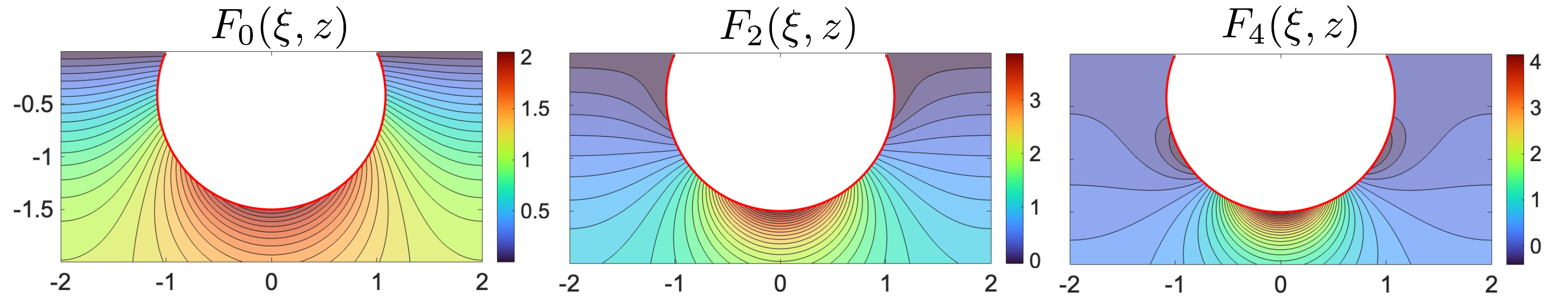}} 
    \caption{$F_0(\xi,z)$, $F_2(\xi,z)$, and $F_4(\xi,z)$ for $\lambda = 1/2$. Using the superposition of these functions, a solution that satisfies (i)--(v) is constructed. }
    \label{fig:F0F2F4}
\end{figure}  

We start the formulation from the solution $\phi_0(\alpha,\beta):=f_\theta(\alpha,\beta)$ defined in~(\ref{eq:pressure_asymp1}). 
Note that this function is harmonic and satisfies $
\phi_0(\alpha,\beta)=0$ for $\beta=2\pi$. Similar to the definition of a new candidate function in~(\ref{eq:phi_2n}), we define a harmonic function that is related to the $2n$-th derivatives of $\phi_0(\alpha,\beta)$ with respect to $\beta$, given by
\begin{align}
    \phi_{2n}(\alpha,\beta) &:= \frac{2\sqrt{2(\cosh \alpha - \cos\beta)}}{(2n)!}\frac{\partial^{2n}}{\partial \beta^{2n}} \int_{0}^\infty \tau \frac{\sinh \theta \tau\sinh(2\pi-\beta)\tau}{\cosh \pi \tau \sinh (\pi-\theta)\tau}P_{-1/2+\rmi\tau}(\cosh\alpha)\d \tau\nonumber\\
    &=\frac{2\sqrt{2(\cosh \alpha - \cos\beta)}}{(2n)!} \int_{0}^\infty \tau^{2n+1} \frac{\sinh \theta \tau\sinh(2\pi-\beta)\tau}{\cosh \pi \tau \sinh (\pi-\theta)\tau}P_{-1/2+\rmi\tau}(\cosh\alpha)\d \tau,
\end{align}
for $n=0,1,\ldots$. 
These functions are harmonic in the domain $D$ and $\phi_{2n}(\alpha,2\pi)=0$, that is, $\phi_{2n}(\alpha,\beta)=0$ for $z=0$, $\xi\geq 1$ in a cylindrical coordinate system. Using the new function $\phi_{2n}(\alpha,\beta)$ and a correction function $\psi_{2n}(\alpha,\beta)$ defined in~(\ref{eq:psi}) with a different kernel,
\begin{align}
    \eta_{2n}(m) = -\frac{2}{\pi}\int_{-\infty}^0 \frac{\sin mz}{mI_1(m/\lambda)}\frac{\partial \phi_{2n}}{\partial \xi}(\alpha(1/\lambda,z),\beta(1/\lambda,z)) \d z,
\end{align}
the solution for $\Pi(\xi,z)$ is given by the superposition of $F_{2n}(\xi,z) := \phi_{2n}(\alpha(\xi,z),\beta(\xi,z)) + \psi_{2n}(\alpha(\xi,z),\beta(\xi,z))$ as follows:
\begin{align}
    \Pi(\xi,z) = -z - \sum_{n=0}^\infty A_{2n} F_{2n}(\xi,z).\label{eq:series_solution1}
\end{align}
This expression is similar to (\ref{eq:ant_semianaly}) by Antkowiak {\em et al.}, except for the different basis functions $F_{2n}(\xi,z)$. The coefficients $A_{2n}$ are determined so that $\Pi(\xi,z)=0$ on $\mathcal{S}$. After obtaining the coefficients $A_{2n}$ using a collocation method, it is possible to calculate the jet speed. This can be done by considering a derivative with respect to $z$ and the chain rule.

Figure~\ref{fig:F0F2F4} shows a contour plot of $F_0(\xi,z)$, $F_{2}(\xi,z)$, and $F_{4}(\xi,z)$ for $\lambda = 0.5$ and $H=1.5$. It can be seen that the higher-order oscillating terms can be included by adding $F_{2n}(\xi,z)$ with a large $n$. Through a linear combination of $F_{2n}$, it is possible to extract the solution $\Pi$ for the boundary value problem~(\ref{eq:problem}). 

\subsection{A semi-analytical solution for a submerged tube}
The same approach can be used to obtain the solution for the problem~(\ref{eq:problem}) with $h\neq 0$. 
First, note that $g_\theta(\alpha,\beta)$ defined in~(\ref{eq:pressure_asymp2}) does not satisfy the boundary condition on the side of the container. To satisfy the boundary condition, 
we consider a correction term $\tilde{\phi}_0(\alpha,\beta):= g_\theta(\alpha,\beta)$, and define
\begin{align}
    G(\xi,z) := \tilde{\phi}_0(\alpha(\xi,z),\beta(\xi,z)) + \tilde{\psi}(\alpha(\xi,z),\beta(\xi,z))
\end{align}
where $\tilde{\psi}(\xi,z)$ is given by~(\ref{eq:psi}) in terms of a Fourier--sine series of $\tilde{\phi}_0(\xi,z)$. Using that, the pressure field is given by
\begin{align}
    \Pi(\xi,z) = -z  -  \sum_{n=0}^\infty \tilde{A}_{2n} F_{2n}(\xi,z)  + h\left(1-G(\xi,z)\right), \label{eq:series_solution2}
\end{align}
where the final term captures the boundary condition at $z=0$ and singularities at $z=0$, $\xi=\pm 1$. After finding the coefficients $\tilde{A}_{2n}$, it is possible to evaluate the field and velocity of the jets.

\begin{figure}
    \centerline{\includegraphics[width=0.99\linewidth]{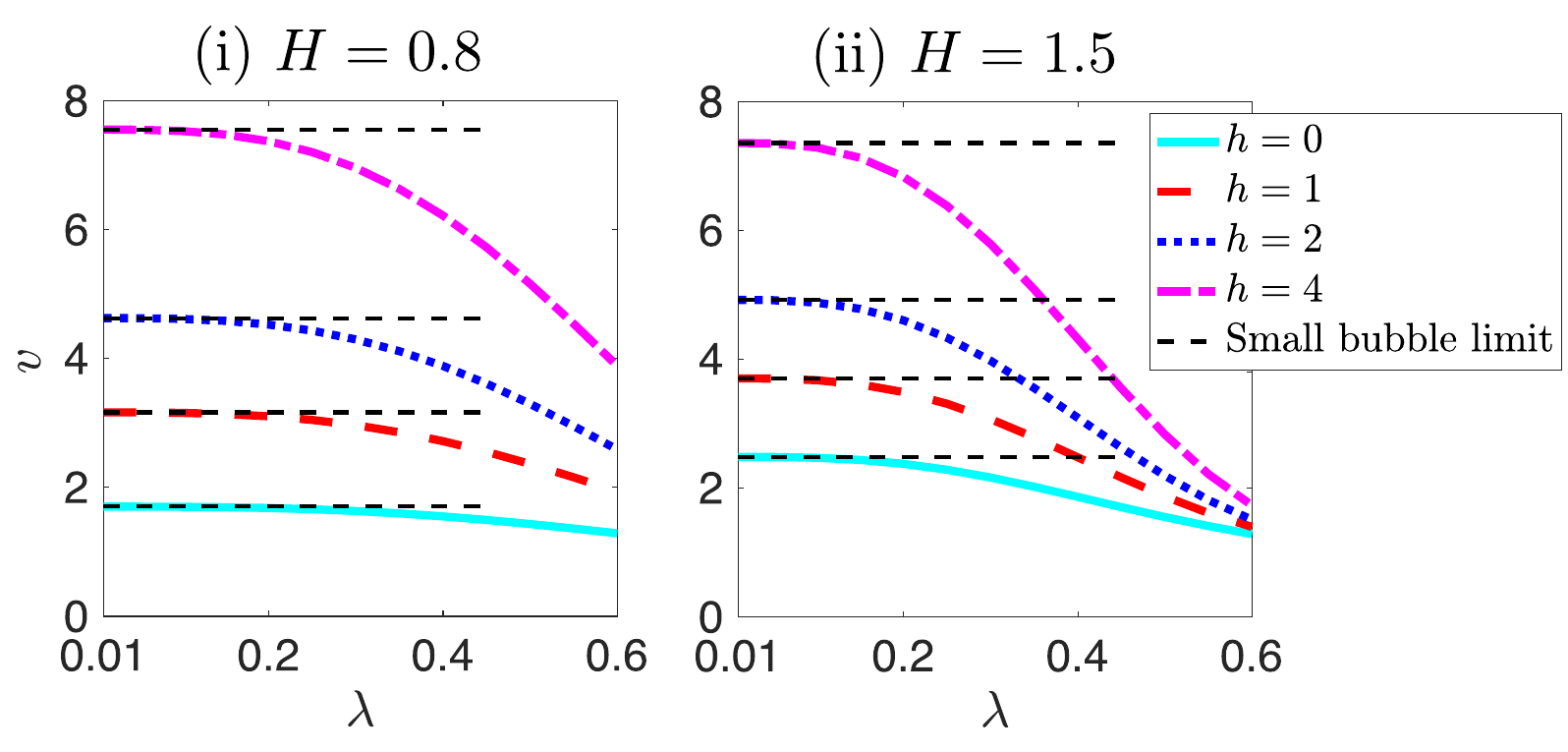}} 
        \caption{A comparison between the analytical solutions~(\ref{eq:pressure_asymp1}) and~(\ref{eq:pressure_asymp2}) for the small-bubble limit plotted as a black dotted line and the numerical solutions calculated as a series of basis functions. 
        (i) $H=0.8$ and (ii) $H=1.5$ with $h=0$, $h=1$, $h=2$, and $h=4$. It can be seen that the accuracy of the analytical solution of the small-bubble limit becomes good as $\lambda$ and $h$ decrease. 
        }
        \label{fig:lambda01}
\end{figure}

For the numerical computation, we truncate the series~(\ref{eq:series_solution1}) at $N$ to calculate the velocity of jets $v(\theta)$ at the bottom of the bubble. Due to the suitability of the basis in the mixed boundary value problem, the required number of basis functions is small. 
Similarly to~\cite{Antkowiak2007-dk}, we find that around 5--10 bases are sufficient for condition (iv), and an absolute error of $10^{-4}$ is obtained. 
In this paper, eight basis functions are used in all computations. The same number of collocation points is chosen on half of $\mathcal{S}$ to construct the linear system for $A_{2n}$ or $\tilde{A}_{2n}$ to satisfy the boundary condition (ii) in~(\ref{eq:F_problem}). 

Figure~\ref{fig:lambda01} compares the analytical solutions~(\ref{eq:velocity1}) and (\ref{eq:velocity2}) for the small-bubble limit, plotted as a black dotted line, and the numerical solutions approximated as a series of basis functions. 
The accuracy of the analytical solution for the small-bubble limit improves as $\lambda$ and $h$ decrease. 
These numerical results indicate that the approximation of the analytical solution derived in~(\ref{eq:velocity1}) and~(\ref{eq:velocity2}) works well for a wide range of $\lambda$ and $h$, especially for small $\lambda$ and $h$.

\begin{figure}
\centerline{\includegraphics[width=0.59\linewidth]{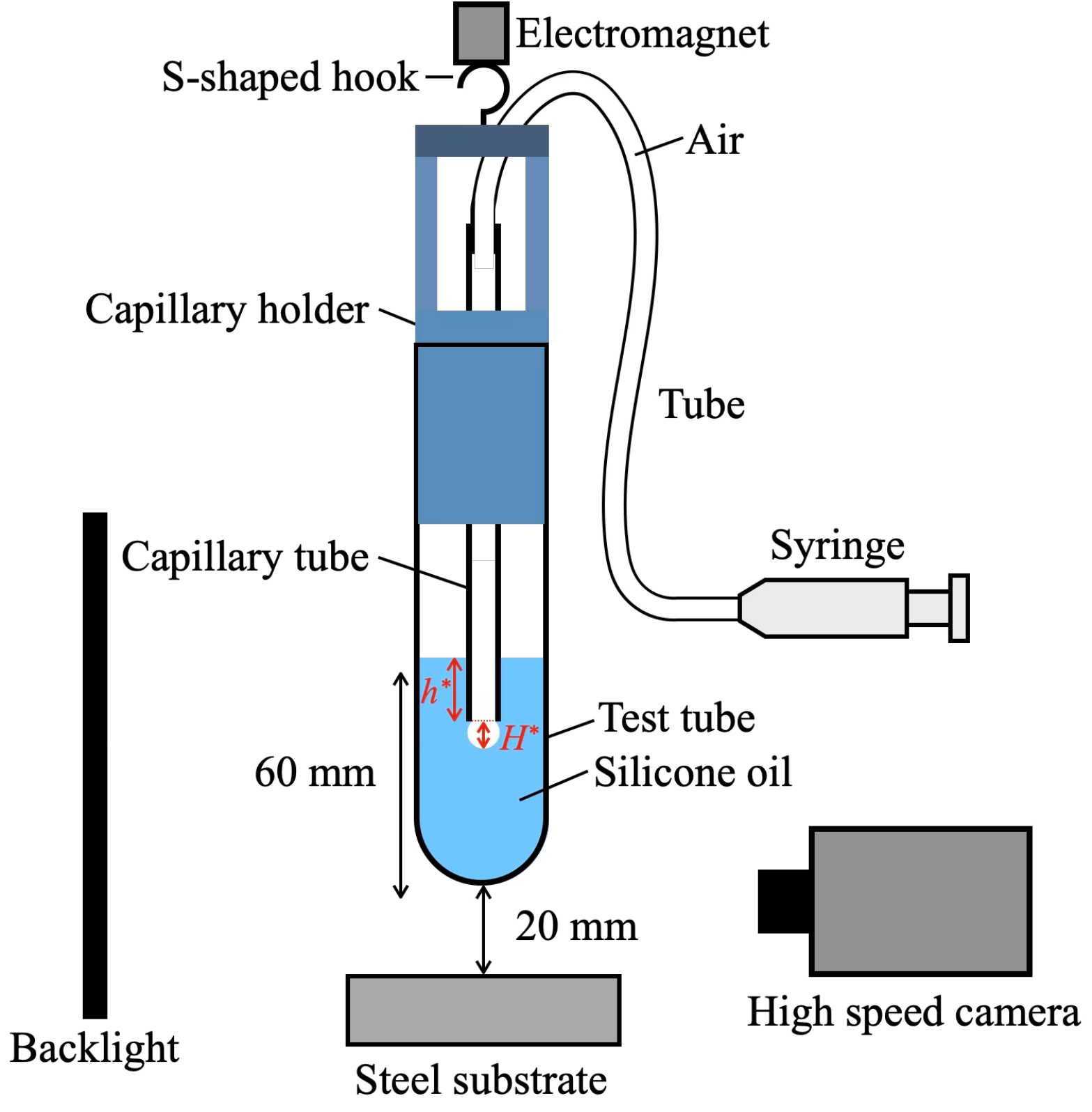}} 
        \caption{A schematic of the experimental setup. A container with an inserted capillary tube is fixed above a steel substrate using an electromagnet, and the container is released by switching off the electromagnet to impact the steel substrate. The depth of the bubble $H^*$ is varied by adjusting the pressure inside the capillary tube using a syringe. The bubble and the jet are recorded using a high-speed camera.
        }
        \label{fig:fig_exp}
\end{figure}

\section{Experiment}\label{sec:experiment}

This section describes the experimental setup adopted in this work and compares the experimental results with analytical formulas. 
The trends of the experimental results are approximated well by the analytical results; that is, there exists an optimal value to produce high velocity jets, as shown in Figure~\ref{fig:figure_vel_maximum}. 

\subsection{Experimental setup}
To produce liquid jets, we followed the example of previous works by~\cite{Kiyama2016-mi,Onuki2018-jx}, 
who studied impulse-driven liquid jets produced in a free-falling container. 
A test tube (IWAKI, 30$\times$200) with an inner diameter of $r^*=13.5$ mm was filled with silicone oil of kinematic viscosity 10 ${\rm mm}^2/{\rm s}$. 
A glass capillary tube with an inner diameter of $R^*=1$ mm was inserted into the test tube and firmly fixed using a capillary holder fabricated with a 3D printer (Bambu Lab, Bambu Lab A1 3D Printer). 
A syringe was used to connect the container to a small tube, then the bubble was injected into the top of the tube. 
A steel S-shaped hook was attached to the top of the capillary holder, and the container was suspended above a steel base plate using an electromagnet. 
The distance between the base plate and the bottom of the container was 20 mm.
When the electromagnet was switched off, the container underwent free fall and collided with the base plate. 
Upon impact, the liquid in the container was rapidly accelerated, ejecting a focused jet from an air bubble at the lower end of the capillary.
A high-speed camera was used to record the behaviour of the bubble during the free fall of the container, as well as the jet formation process.
The parameter $\lambda$, which is the ratio of the radius of the container outside the tube to the radius of the tube, was $0.074$ in all experiments. For this $\lambda$, it was possible to approximate the jet velocity using the analytical formulas~(\ref{eq:velocity1}) and (\ref{eq:velocity2}), as shown in Figure~\ref{fig:lambda01}. 
The depth of the viscous liquid was set to $60$ mm, which was assumed to be much larger than the tube radius. 

The experimental parameters were $h^*$ and $H^*$. 
The height $H^*$ of the tube tip was measured using a horizontally placed camera as the tube fell. Three values of $h^*$ were examined: $h^*=0$ mm, $h^*=2$ mm, and $h^*=4$ mm.  
As $H^*$ oscillated while the tube was falling, a video was captured before and after impact to determine $H^*$ immediately before impact by visual inspection. 
The position of the jet tip generated just after impact was also identified visually, and the jet velocity was then calculated. 
Experiments were conducted to measure the jet speed as a function of the bubble height $H^*$ and immersion depth $h^*$. 

\subsection{A comparison of the analytical solution with experimental results}

\begin{figure}
    \centerline{\includegraphics[width=0.99\linewidth]{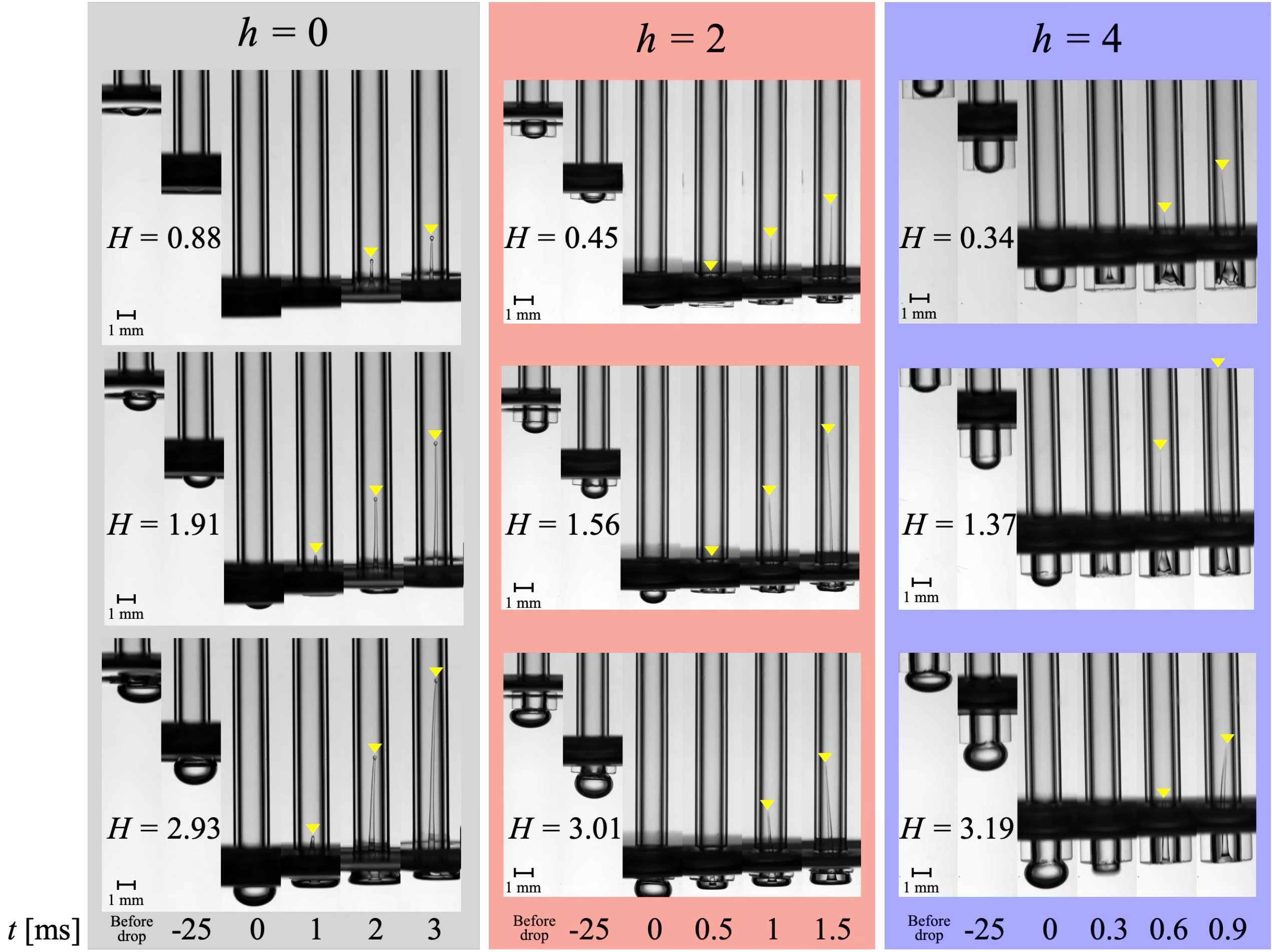}} 
        \caption{Image sequences of the jet. The moment of impact between the container and the substrate is defined as $t = 0$ ms. For each condition, six images are shown, including images taken before the container is released and 25 ms before impact. The yellow triangles indicate the jet tip position. The bubble grows as the container falls and becomes nearly spherical. For $h = 0$, the jet velocity increases with increasing $H$. In contrast, for $h > 0$, there exists a value of $H$ at which the jet velocity reaches a maximum, and this value depends on $h$.
For $h = 2$, the jet velocity is maximized at $H = 1.56$, while for $h = 4$, it reaches a maximum at $H = 1.37$.  }
        \label{fig:jet_exp}
\end{figure}  

Figure~\ref{fig:jet_exp} shows image sequences of the jet with different $h=h^*/R^*$ and $H=H^*/R^*$. We set $t=0$ as the time when the container collided with the base. The jet formation process is shown at time intervals of 1 ms for $h=0$, 0.5 ms for $h=2$, and 0.3 ms for $h=4$. 
For each condition, the image sequences in Figure~\ref{fig:jet_exp} also include the spherical bubble shape before the container is released, as well as that during the free fall of the container (25 ms before impact). 
For $h=0$ and $h=2$, the bubbles are partially hidden by the interface, making their shapes difficult to observe. Under all conditions, the bubble approaches a nearly spherical, laterally symmetric shape during the free fall of the container. 
Subsequently, when the container impacts the base plate, the bubble undergoes significant deformation, leading to the ejection of a focused jet. 
Focusing on the jet tip position at $t=3$ ms for $h=0$, it is observed that the jet tip height increases with increasing $H$. 
This indicates that, for 
$h=0$, the jet velocity increases as 
$H$ increases. 
In contrast, for $h>0$, a larger value of 
$H$ does not necessarily lead to a higher jet tip position. 
For $h=2$, the jet tip position at 
$t=1.5$ ms shows that the maximum jet height is obtained at 
$H=1.56$. Similarly, for 
$h=4$, the jet tip position at 
$t=0.9$ ms reaches its maximum at 
$H=1.37$. These results suggest that, for 
$h>0$, there exists an optimal value of $H$ at which the jet velocity is maximized, and that this optimal value depends on $h$. 

To fit our analytical solution with the experimental data, 
we first non-dimensionalized the measured velocity $U_\theta^*(h)$ using the velocity of the container $U_0^*$ just after impact as follows: 
\begin{align}
    U_{\theta}(h) := \frac{U_{\theta}^*(h)- U_0^*}{U_0^*}.
\end{align}
To compare the analytical solutions with the experimental data, we considered the following scaling factor $\gamma(h)$, which depends only on the height $h$: 
\begin{align}
    U_{\theta}(h) \approx \gamma(h) (v_f(\theta) + h v_g(\theta)).
\end{align}
The scaling factor $\gamma(h)$ is needed because there may exist some scaling differences due to the experimental setup~\citep{Kiyama2016-mi}.  
The value $\gamma(h)$ was determined through fitting of the experimental data using the least squares method. 



\begin{figure}
    \centerline{\includegraphics[width=0.99\linewidth]{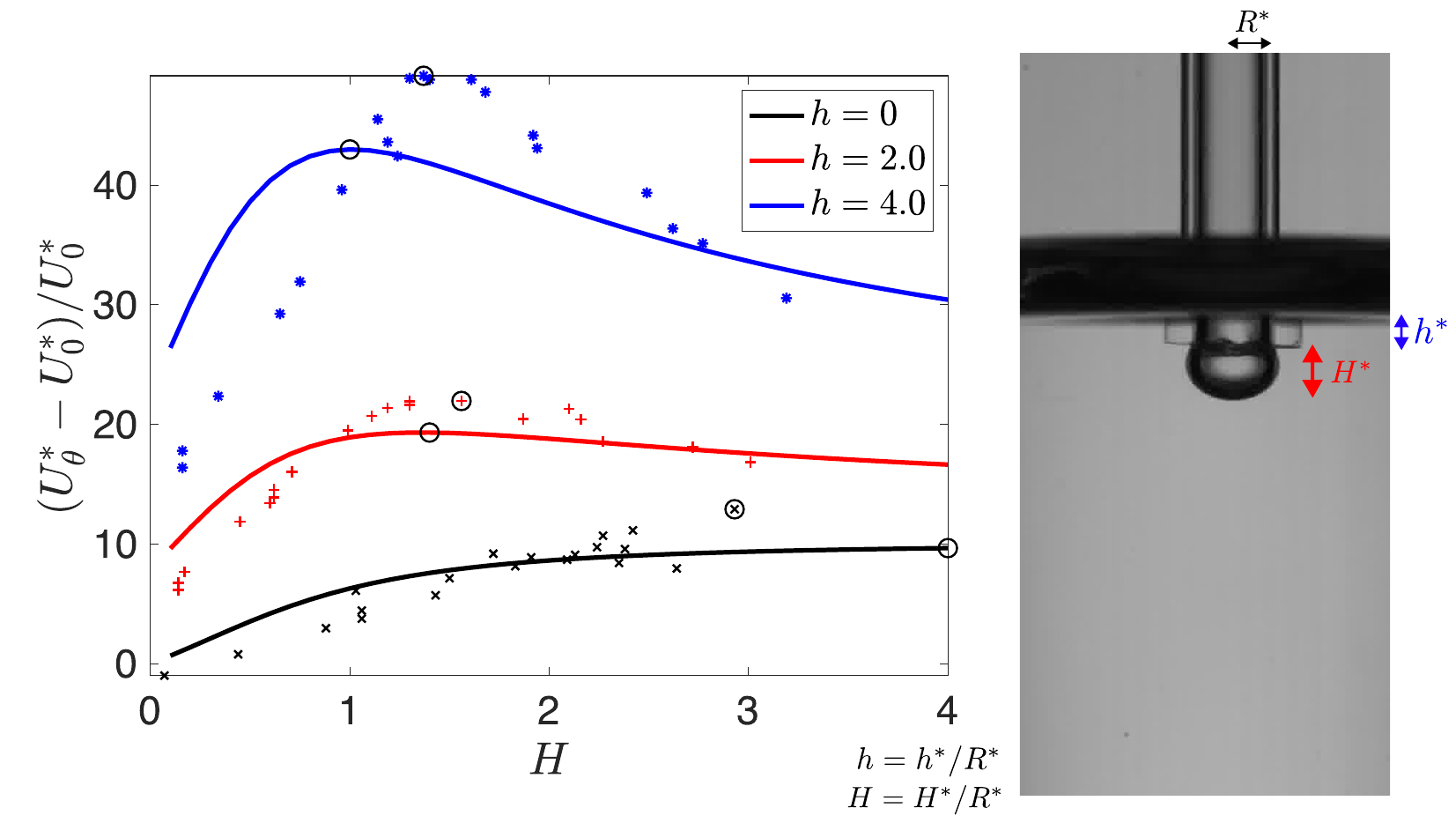}} 
        \caption{A comparison between the analytical solution for the small-bubble limit in Eqs. (\ref{eq:pressure_asymp1}) and (\ref{eq:pressure_asymp2}), plotted as solid lines, and the experimental data, plotted as dots. 
        The data are plotted for $h=0$, $h=2.0$, and $h=4.0$. 
        Black circles correspond to the maximum jet velocity in both our theoretical and experimental results.  
        }
        \label{fig:exp}
\end{figure}

Figure~\ref{fig:exp} compares the analytical solution for the small-bubble limit in Eqs. (\ref{eq:pressure_asymp1}) and (\ref{eq:pressure_asymp2}), 
plotted as solid lines, and the experimental data, plotted as dots. 
Both experimental and analytical data are plotted for $h=0$, $h=2.0$, and $h=4.0$. 
Each black circle corresponds to a maximum jet velocity in our experiments. 
Although a scaling factor $\gamma(h)$ is needed to fit the experimental data, 
the trend of the experimental data is approximated well by the analytical solution $v(\theta)$.  
This indicates that controlling the bubble height $H$ and the immersion depth $h$ is important for producing a high-velocity jet.

\section{Discussions}\label{sec:conclusion}

This paper has derived an analytical solution for the pressure impulse outside a spherical bubble, generalizing previous work by~\cite{Antkowiak2007-dk}. 
The jet speed was calculated by simple integration of hyperbolic functions, indicating that there exists a local maximum jet velocity for a specific bubble height $H$ and immersion depth $h$. 

\begin{figure}
    \centerline{\includegraphics[width=0.99\linewidth]{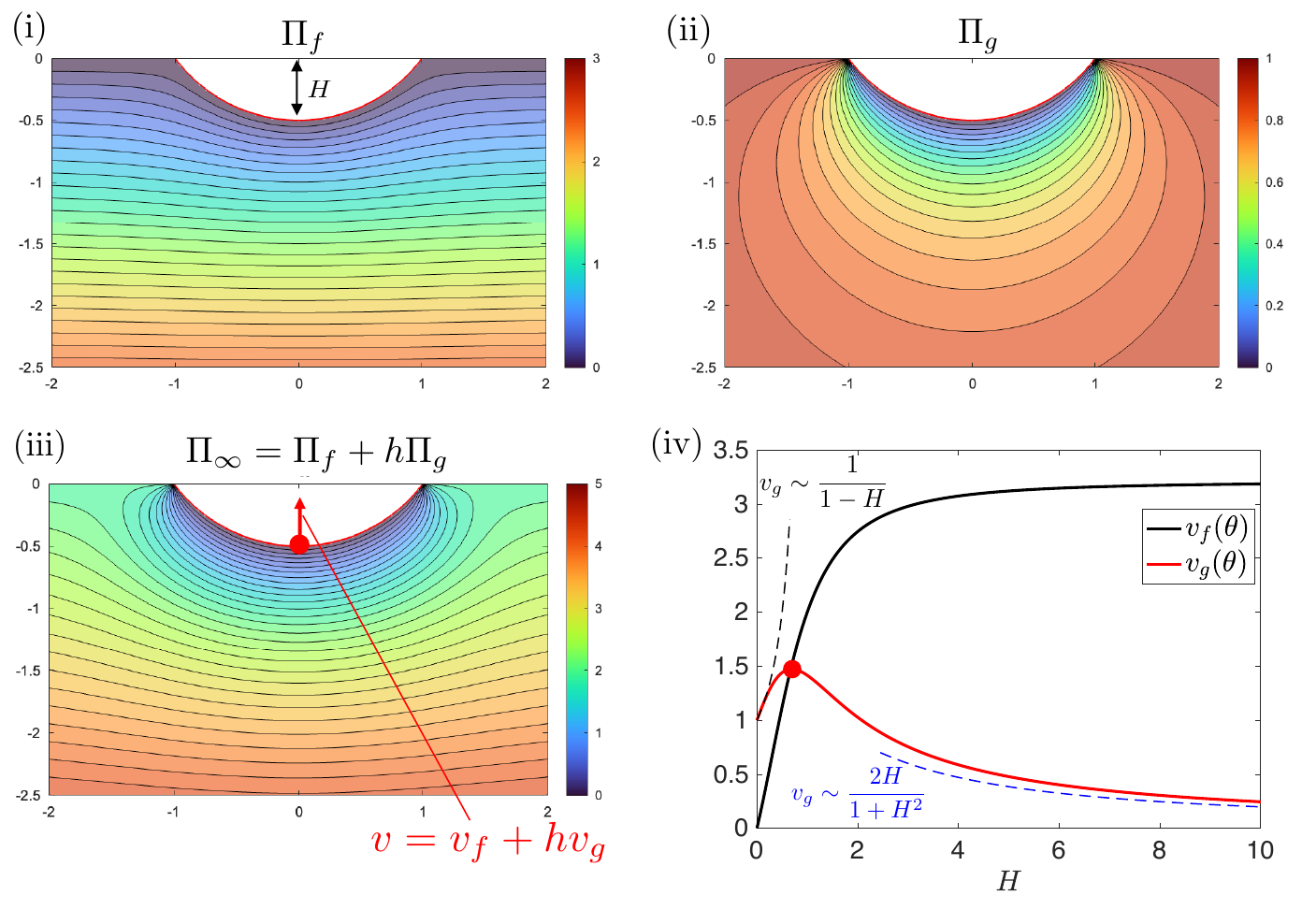}} 
        \caption{A visual explanation of why the velocity $v(\theta)$ has an optimal value for a specific depth $h$ and bubble height $H$. The pressure impulse with a submersion depth $h$ is $\Pi_f + \Pi_g \times h$. The velocity $v_f$ induced by $\Pi_f$ is a monotonically increasing function with respect to $H$, but the $v_g$ induced by $\Pi_g$ attains its maximum value around $H=0.7$. Thus, there exists an optimal value for $H$ that maximizes the velocity $v(\theta) = v_f(\theta)+hv_g(\theta)$ with respect to $h$. }
        \label{fig:jet_comp}
\end{figure}  

It should be noted that there is an optimal bubble shape for a submerged tube in a container that produces the fastest jet. This can be explained physically. The total velocity $v(\theta) = v_f(\theta) + h v_g(\theta)$, where $v_f(\theta)$ is a velocity induced by the pressure impulse with $h=0$ shown in (i) in Figure~\ref{fig:jet_comp}, and $v_g(\theta)$ is a velocity due to the height of the water in the container in (ii) in Figure~\ref{fig:jet_comp}. The velocity $v_f(\theta)$ is a monotonically increasing function, because the equi-contour line of $\Pi_f$ is increasingly distorted as the bubble size increases, as shown in Figure~\ref{fig:jet_comp}. This produces the large pressure gradient on the bottom of the bubble, and hence the largest jet velocity. 

For $v_g(\theta)$, however, our numerical calculation shows that there exists a local maximum of $v_g(\theta)$, as shown in Figure~\ref{fig:jet_comp}. 
This can be explained through the analogy of electrical potential. The velocity $v_g(\theta)$ is given by $-\partial g_\theta/\partial z$ with~(\ref{eq:pressure_asymp2}) evaluated at the bottom. 
As $H$ becomes large, the bubble can be seen as a sphere with a radius of $1/\sin(\theta) \rightarrow \infty$ and a potential of $0$, attached to the ground with a potential of $1$. This gives us the asymptotic electrical potential as 
\begin{align}
g_\theta(\alpha(\xi,z),\beta(\xi,z)) \sim 1-\frac{\sin \theta}{\sqrt{\xi^2 + (z- 1/\tan \theta)^2}},
\end{align}
which gives
\begin{align}
    v_g(\theta) = -\frac{\partial g_\theta}{\partial z}(0,\pi+\theta) \sim \frac{2H}{1+H^2}, \label{eq:asymp1_vg} 
\end{align}
where we used~(\ref{eq:H_theta}). This gives the asymptotic behaviour of $v_g(\theta)$ as $H\rightarrow \infty$. 

For small $H$, the distance from the top of the bubble to the ground region $z=0$ gets small as $H$ increases, since the distance is measured as the curved line along electric flux lines. By considering $v_g(0)=1$ and defining the virtual distance from the bottom of the bubble to the ground as $r_v$, the velocity $v_g$ is assumed to have an asymptotic behaviour as follows:
\begin{align}
    v_g (\theta) \sim \frac{r_v}{r_v - H} = \frac{1}{1 - H/r_v}. \label{eq:asymp2_vg}
\end{align}
Figure~\ref{fig:jet_comp} shows (iv) two asymptotic behaviours~(\ref{eq:asymp1_vg}) and (\ref{eq:asymp2_vg}) with $r_v=1$ compared with our numerical calculation. 
These two asymptotics agree well with our numerical calculation of $v_g$. Moreover, these behaviours show that there should exist a local maximum for $v_g(\theta)$. This is one explanation for why there exists an optimal shape for producing the fastest jet. 

Although our approach works for small $\lambda$, a correction term for the analytical solutions~(\ref{eq:pressure_asymp1}) and (\ref{eq:pressure_asymp2}) is necessary for 
deriving the jet speed when $\lambda$ is greater than approximately $0.3$. 
In that case, matched asymptotic expansions for the pressure impulse at the outer boundary in the small-bubble limit could provide a promising approach. 
In general, matched asymptotic expansions match the behaviour of the inner region to that of the outer region, as initially studied by~\cite{van1975perturbation}. A review of general perturbation techniques is described in~\cite{masoud2019reciprocal}.
Recently, this approach has been used in various fields of fluid problems, such as to evaluate the evaporation of multiple bubbles~\citep{Masoud2021-ke} and slip lengths of superhydrophobic channels. Friction resistances are also approximated well by matched asymptotic expansions (see~\cite{Miyoshi2024-heat_sink,Miyoshi2024-sl,Hiroyuki_Miyoshi_Darren_G_Crowdy_undated-mq,Rodriguez-Broadbent2024-kf}). Future work will focus on this topic to elucidate the effect of side walls to derive another closed-form solution instead of a series solution. 


In summary, this work has established an expanded analytical framework for the formation of impulse-driven jets from curved fluid interfaces. By solving the pressure--impulse problem for spherical bubbles with arbitrary contact angles, we have shown that the jet velocity is governed by the competition between curvature-induced impulse focusing and submersion-induced pressure redistribution. This competition gives rise to an optimal interface geometry that maximizes jet speed, providing a clear physical explanation for experimentally observed non-monotonic trends.

Beyond the specific configuration considered here, the present formulation offers a foundation for systematically exploring impulse focusing in more general axisymmetric geometries. 
The ability to relate interface curvature directly to jet velocity opens a pathway to geometry-controlled jet engineering, such as conical interfaces, with potential implications for high-speed microjet generation, flow focusing devices, and impact-driven injection technologies.

\subsection*{Acknowledgement}
The first author is supported by a JSPS Postdoctoral Fellowship (JSPS-24KJ0041) and an Excellent Young Researcher Fellowship at the University of Tokyo. 
This work was supported by JSPS KAKENHI Grant No. JP24H00289. 

\subsection*{Declaration of Interests}
The authors report no conflict of interest.

\appendix

\section{Derivation of the analytical formulas~(\ref{eq:pressure_asymp1}) and (\ref{eq:pressure_asymp2}) }\label{app:1}
This section explains how the 3D Laplace equations can be solved using a toroidal coordinate system. Following the mathematical setup provided on p. 222 of~\cite{lebedev1965special}, we consider a change of variables from cylindrical coordinates $(\xi,z)$ to $(\alpha,\beta)$ as follows:
\begin{align}
      \xi=\frac{\sinh\alpha}{\cosh\alpha-\cos\beta},\quad z=\frac{\sin\beta}{\cosh\alpha-\cos\beta},
\end{align}
where the negative semi-infinite region outside the bubble domain is parameterized by
\begin{align}
  0 \leq \alpha < \infty,\qquad \pi+\theta<\beta \leq 2\pi.
\end{align}
Using these new coordinates, the boundary of the spherical bubble $\mathcal{S}$ corresponds to $\beta=\pi + \theta$, and the region $\xi\geq 1$ corresponds to $\beta = 2\pi$.    
The derivatives of the coordinates $(\alpha,\beta)$ with respect to $\xi$ and $z$ are given by  
\begin{align}
    \frac{\partial \alpha}{\partial \xi} = \frac{\xi+1}{z^2+(1+\xi)^2} - \frac{\xi-1}{z^2+(1-\xi)^2},\ \ \frac{\partial \beta}{\partial \xi} = -\frac{4z\xi}{(\xi^2 + z^2 - 1)^2 + 4z^2},
\end{align}
and
\begin{align}
    &\frac{\partial \alpha}{\partial z} = z\left(\frac{1}{z^2 + (1+\xi)^2} - \frac{1}{z^2 + (1-\xi)^2}\right),\ \frac{\partial \beta}{\partial z} = \frac{2(\xi^2 -z^2-1)}{(\xi^2 + z^2 - 1)^2 + 4z^2}. 
\end{align}
Through direct calculation, we can calculate the metric tensor $h_{\alpha}$, $h_{\beta}$, and $h_{\varphi}$ as follows:
\begin{align}
    h_\alpha = h_\beta = \frac{1}{\cosh\alpha - \cos\beta},\quad h_\varphi = \frac{\sinh\alpha}{\cosh\alpha - \cos\beta},
\end{align}
where $\varphi$ is the azimuthal angle. 
The 3D Laplace equation in cylindrical coordinates with a rotational symmetry condition then becomes
\begin{align}
    \frac{\partial}{\partial \alpha}\left(\frac{\sinh\alpha}{\cosh\alpha - \cos\beta}\frac{\partial u}{\partial \alpha} \right) + \frac{\partial}{\partial \beta}\left(\frac{\sinh\alpha}{\cosh\alpha - \cos\beta}\frac{\partial u}{\partial \beta} \right) = 0.
\end{align}
Considering the change of variables $u=\sqrt{2(\cosh\alpha-\cos\beta)}v$, we get 
\begin{align}
  \frac{\partial^2 v}{\partial \alpha^2} + \frac{\partial^2 v}{\partial \beta^2} + \coth\alpha\frac{\partial v}{\partial \alpha}+\frac{1}{4}v = 0.
\end{align}
Now we can use a separation of variables approach. Setting $v=A(\alpha)B(\beta)$, we have 
\begin{equation}
  \left\{ \,
    \begin{aligned}
    &\frac{\d^2B}{\d\beta^2} = - \nu^2 B\\
    &\frac{1}{\sinh\alpha}\frac{\d}{\d\alpha}(\sinh\alpha\frac{\d A}{\d\alpha}) - \left(\nu^2-\frac{1}{4}+\frac{\mu^2}{\sinh^2\alpha}\right) = 0.
    \end{aligned}
  \right.
\end{equation}
The solution for $u$ can be represented by the superposition of the following functions:
\begin{align}
    u(\alpha,\beta) = \sqrt{2(\cosh \alpha - \cos\beta)} [M_\tau \cosh \beta\tau + N_\tau \sinh \beta \tau]P_{-1/2 + \rmi \tau}(\cosh\alpha),
\end{align}
where the Legendre function $v=P_{\nu}(z)$ satisfies 
\begin{align}
  (1-z^2)v'' - 2zv' + \nu(\nu+1)v = 0.
\end{align}

To derive the formulas for the pressure impulse~(\ref{eq:pressure_asymp1}) defined by $f_\theta(\alpha,\beta)$, 
we follow the procedure in Section 8.12 in~\cite{lebedev1965special}. 
The boundary condition for the function $f_\theta(\alpha,\beta)$ is given as follows:
\begin{align}
    \left\{
    \begin{aligned}
        &f_\theta(\alpha,\beta)=-z = -\frac{\sin \beta }{\cosh\alpha - \cos\beta},\quad {\rm for}\ \alpha\geq 0,\  \beta= \pi+\theta,\ 2\pi.
    \end{aligned}
    \right.\label{eq:f_theta_problem}
\end{align}
Note that the condition of $f_\theta \rightarrow 0$ for $z\rightarrow -\infty$ is automatically satisfied by the boundary condition $f_\theta(\alpha,2\pi) =0$ as $\alpha \rightarrow 0$.     

Now we use the property of the Legendre function $P_{-\frac{1}{2} + \rmi \tau}(\cosh\alpha)$ as follows~\citep{lebedev1965special}:
\begin{align}
    \frac{1}{\sqrt{2\cosh\alpha - 2\cos\beta}} = \int_0^\infty \frac{\cosh(\pi-\beta)\tau}{\cosh \pi\tau} \Pmt(\cosh\alpha)\d \tau, \label{eq:sqrt2}
\end{align}
and the derivative of (\ref{eq:sqrt2}) with respect to $\beta$ is given by 
\begin{align}
-\frac{\sin\beta}{(2\cosh\alpha - 2\cos\beta)^{3/2}} = -\int_0^\infty \tau\frac{\sinh(\pi-\beta)\tau}{\cosh \pi\tau}  \Pmt(\cosh\alpha)\d \tau.\label{eq:sqrt2_deriv}
\end{align}
The left-hand side of (\ref{eq:sqrt2_deriv}) is given by 
\begin{align}
    -\frac{\sin\beta}{(2\cosh\alpha - 2\cos\beta)^{3/2}} = 2 \cdot \frac{f_\theta(\alpha,\beta)}{\sqrt{2(\cosh\alpha - \cos\beta)}},\quad {\rm for} \ \beta = \pi +\theta,\ 2\pi,
\end{align}
which gives the coefficients of the integrand in Eq. (8.12.11) in~\cite{lebedev1965special} as follows:
\begin{align}
    \Phi_1 = 0,\quad \Phi_2 = 2 \tau \frac{\sinh \theta \tau}{\cosh \pi \tau}
\end{align}
Now, by substituting the integrand of (\ref{eq:sqrt2_deriv}) into Eq. (8.12.11) with $\beta_1 = 0$, and $\beta_2 = \pi+\theta$, we obtain
\begin{align}
    f_\theta(\alpha,\beta) = 2\sqrt{2(\cosh\alpha - \cos\beta)}\int_0^\infty \tau \frac{\sinh \theta \tau \sinh(2\pi-\beta)\tau }{\cosh \pi\tau \sinh (\pi-\theta)\tau }P_{-\frac{1}{2} + \rmi\tau}(\cosh\alpha)\d \tau,
    \label{eq:pressure_asympt1}
\end{align}
Similar manipulation gives the formula for $g_\theta(\alpha,\beta)$ defined in~(\ref{eq:pressure_asymp2}). 
Note that the function $g_\theta(\alpha,\beta)$ is derived using the same procedure as in the example on p. 229 of~\cite{lebedev1965special}, where a boundary condition that has a constant value on the spherical cap is considered.

Note that $\partial \alpha/\partial z = 0$ and 
\begin{align}
    \frac{\partial \beta}{\partial z}(0,-H) = \frac{-2(H^2+1)}{(H^2-1)^2 + 4H^2} = -\frac{\sin^2\theta}{1-\cos\theta}.
\end{align}
The derivative with respect to $\beta$ is
\begin{align}
    \frac{\partial f_\theta}{\partial \beta}(\alpha,\beta) &= \frac{2\sin \beta}{ \sqrt{2(\cosh\alpha - \cos\beta)}}\int_{0}^\infty \tau \frac{\sinh \theta \tau \sinh(2\pi-\beta)\tau }{\cosh \pi\tau \sinh (\pi-\theta)\tau }P_{-\frac{1}{2} + \rmi\tau}(\cosh\alpha)\d \tau\nonumber\\
    &-2\sqrt{2(\cosh\alpha - \cos\beta)}\int_0^\infty \tau^2 \frac{\sinh \theta \tau \cosh(2\pi-\beta)\tau }{\cosh \pi\tau \sinh (\pi-\theta)\tau }P_{-\frac{1}{2} + \rmi\tau}(\cosh\alpha)\d \tau.
\end{align}
Using $P_{-\frac{1}{2}+\rmi \tau}(1) = 1$, we arrive at     
\begin{align}
    \frac{\partial f_\theta}{\partial \beta}(0,\pi+\theta) = \sqrt{\frac{2}{1 + \cos\theta}}\int_{0}^\infty\tau\frac{\sinh\theta\tau}{\cosh\pi \tau}\left(\sin\theta + 2(1+\cos\theta)\tau\frac{\cosh(\pi-\theta)\tau}{    \sinh(\pi-\theta)\tau} \right)\d\tau.
\end{align}
Then, using some identities of Legendre function $\Pmt$ and its derivative derived from the Mehler--Fock theorem, 
\begin{align}
    \frac{\partial f_\theta}{\partial z}(0,\pi+\theta) = \sin^2 \left(\frac{\theta}{2} \right) + 8 \cos^3\left(\frac{\theta}{2}\right)\int_0^{\infty} \tau^2 \frac{\sinh\theta \tau \cosh(\pi-\theta)\tau}{\cosh\pi\tau \sinh(\pi-\theta)\tau}\d \tau.
\end{align}
The derivative of $g_\theta$ with respect to $z$ is obtained in a similar manner. 

\section{Computation of the Legendre Function $P_{-\frac{1}{2}+\rmi \tau}(\cosh\alpha)$}\label{app:comp_legendre}
A standard MATLAB platform was used for all numerical computations. However, the original ``legendreP'' function in Matlab does not accept a complex number for the order. It was thus necessary to implement a numerical code to compute $P_{-\frac{1}{2}+\rmi \tau}$. 

It is known that $P_{-\frac{1}{2} + \rmi \tau}(\cosh\alpha)$ has some integral representations: 
\begin{align}
    P_{-\frac{1}{2}+\rmi \tau}(\cosh\alpha) = \frac{2}{\pi}\cosh \pi\tau \int_0^\infty \frac{\cos \tau\theta}{\sqrt{2\cosh\theta + 2\cosh\alpha}} \d \theta.\label{eq:Pmt_1}
\end{align}
We observe that the representation~(\ref{eq:Pmt_1}) works numerically well for the range of $\tau < \tau_0\in \mathbb{R}$. 
Although $P_{-\frac{1}{2}+\rmi \tau}(\cosh\alpha)$ is convergent as $\tau\rightarrow \infty$, the representation~(\ref{eq:Pmt_1}) becomes numerically unstable due to the term $\cosh\pi \tau$ for large $\tau$. 
For large $\tau$, we use the representation~(\ref{eq:Pmt_2}) instead of~(\ref{eq:Pmt_1}):
\begin{align}
    P_{-\frac{1}{2}+\rmi \tau}(\cosh\alpha) = \frac{2}{\pi \tanh \pi \tau}\int_\alpha^\infty \frac{\sin \tau\theta}{\sqrt{2\cosh\theta - 2\cosh\alpha}} \d \theta. \label{eq:Pmt_2}
\end{align}
The derivative of $\Pmt(\cosh\alpha)$ with respect to  $\alpha$ is calculated using the derivative of the integrand of Eq.~(\ref{eq:Pmt_1}). 
For all numerical calculations, we use a fixed value $\tau_0 = 3$. 
Note that, for $\alpha=0$, it is easy to see that $P_{-\frac{1}{2}+\rmi \tau}(1) = 0$.



\bibliographystyle{jfm}
\bibliography{references}

\end{document}